\RequirePackage{fix-cm}
\documentclass[smallextended]{svjour3}       
\smartqed  
\usepackage{graphicx}

\usepackage{amsmath}
\usepackage{amssymb}
\usepackage{placeins}

\begin{document}

\title{Algorithmic simulation of far-from-equilibrium dynamics using quantum computer
\thanks{W. V. P. acknowledges a support from RFBR (project no. 15-02-02128). Yu. E. L. acknowledges a support from RFBR (project no. 17-02-01134) and the Program of Basic Research of HSE.}
}

\titlerunning{Algorithmic simulation of far-from-equilibrium dynamics}        

\author{A.\,A.~Zhukov         \and
    S.\,V.~Remizov \and
    W.\,V.~Pogosov \and
    Yu.\,E.~Lozovik 
}


\institute{A.\,A.~Zhukov \at
              Dukhov Research Institute of Automatics (VNIIA), 127055 Moscow, Russia \\
              National Research Nuclear University (MEPhI), 115409 Moscow, Russia
           \and
    S.\,V.~Remizov  \at
    Dukhov Research Institute of Automatics (VNIIA), 127055 Moscow, Russia \\
    Kotel'nikov Institute of Radio Engineering and Electronics, Russian Academy of Sciences, 125009 Moscow, Russia
    \and
    W.\,V.~Pogosov   \at
    Dukhov Research Institute of Automatics (VNIIA), 127055 Moscow, Russia \\
    Institute for Theoretical and Applied Electrodynamics, Russian Academy of
    Sciences, 125412 Moscow, Russia\\
    Tel.: +7(926)359-6034\\
    Fax.: +7(499)978-0903\\
    \email{Walter.Pogosov@gmail.com}           
    \and
    Yu.\,E.~Lozovik \at
    Dukhov Research Institute of Automatics (VNIIA), 127055 Moscow, Russia \\
    Institute of Spectroscopy, Russian Academy of Sciences, 142190 Moscow, Russia\\
    Moscow Institute of Electronics and Mathematics, National Research University Higher School of Economics, 101000 Moscow, Russia
}

\date{Received: date / Accepted: date}

\maketitle

\begin{abstract}
We point out that superconducting quantum computers are prospective for the simulation of the dynamics of spin models far from equilibrium, including nonadiabatic phenomena and quenches. The important advantage of these machines is that they are programmable, so that different spin models can be simulated in the same chip, as well as various initial states can be encoded into it in a controllable way. This opens an opportunity to use superconducting quantum computers in studies of fundamental problems of statistical physics such as the absence or presence of thermalization in the free evolution of a closed quantum system depending on the choice of the initial state as well as on the integrability of the model. In the present paper, we performed proof-of-principle digital simulations of two spin models, which are the central spin model and the transverse-field Ising model, using 5- and 16-qubit superconducting quantum computers of the IBM Quantum Experience. We found that these devices are able to reproduce some important consequences of the symmetry of the initial state for the system's subsequent dynamics, such as the excitation blockade. However, lengths of algorithms are currently limited due to quantum gate errors. We also discuss some heuristic methods which can be used to extract valuable information from the imperfect experimental data.


\end{abstract}

\section{Introduction}

Quantum computers and simulators are prospective for the resolution of problems which are hard to solve using conventional computing systems. In principle, these quantum devices can be constructed on the basis of different physical platforms, see, e.g., Refs. \cite{Girvin,Blatt,Lukin,Harty,Lu,Maller,Zwan,Nori}. However, over last years, superconducting realization seemed to become most promising for the construction of large-scale programmable quantum computers. Quantum processors of several types have been created and various algorithms have been implemented to show concepts of error correction \cite{Matrinis1,DiCarlo,Gambetta,Chow}, modeling spectra of molecules \cite{variat} and other fermionic systems \cite{Hubbard}, simulation of light-matter systems \cite{LM}, many-body localization \cite{MBL}, machine learning \cite{ML}, scaling issues \cite{Rigetti} etc. Besides, superconducting quantum circuits provide a unique platform to study the effects of quantum optics and nonstationary quantum electrodynamics, see, e.g., Refs. \cite{Devoret,Astafiev,Macha,Oelsner,DCE1,DCE2,Segev,We1,We2,We3}.

State-of-the-art processors contain tens of superconducting qubits with individual control and readout. Quantum computer of such a size might enable for the first time to demonstrate advantages over modern and most capacitive conventional supercomputers \cite{supremacy1,Chow}. Indeed, the dimension of Hilbert space needed to store a highly entangled state of 50 qubits is $2^{50}$. Storage and manipulation of such a state is beyond the capabilities of best modern supercomputers.

In the present paper, we point out that superconducting quantum computers may provide a new platform to study far-from-equilibrium dynamics of various spin models in one and two dimensions. Spin models are nowadays considered as a playground to study fundamental aspect of statistical physics such as thermalization of closed quantum systems during their free evolution from an initial state differing from the Hamiltonian eigenstate, see, e.g., Refs. \cite{Calabrese,Rieger,Eisert,Gogolin1,Polkovnikov,Polkovnikov1}. Recently there was an impressive progress in trapped cold atomic gases which can serve as analog simulators of spin models and boosted theoretical research in this field \cite{Hofferberth,Trotzky,Rigol,Bloch,Gogolin1,Polkovnikov2}. The lack of thermalization observed experimentally in quasi-one-dimensional geometry has been attributed to the integrability of underlying model. Later on, the dynamical behavior was argued to be also dependent on the initial state \cite{Banuls,Gogolin,Polkovnikov,Polkovnikov1}. Programmable quantum computers have an advantage that the dynamics of different spin models can be modeled via unitary evolution using the same chip. Another positive aspect is that various initial conditions including entangled states of spins can be implemented thanks to the individual addressability of physical qubits in quantum computers. Therefore, it is also possible to analyze the dependence of the evolution on the initial conditions. Note that spin models are also directly applicable to study magnetic properties of various materials, see, e.g., Ref. \cite{Rosner}, so that quantum simulation of these models can have an impact on material science as well.

We here perform proof-of-principle simulation of the far-from-equilibrium dynamics in two different spin models using 5-qubit and 16-qubits IBM quantum computers, which are available through the internet within IBM Quantum Experience. The models we study are the central spin model and the transverse-field Ising model the latter being both in the one-dimensional and ladder configurations. The choice of these two models is optimal from the viewpoint of the topologies of available chips as well as existing errors of two-qubit gates. Unitary evolution of the system from initial states, which are not exact eigenstates of systems Hamiltonians, is implemented in a digital way using Trotter decomposition of the evolution operator.  We demonstrate that modern quantum computers are able to reproduce important aspects of spin dynamics originating from the symmetry of the initial entangled state, such as the excitation blockade due to the quantum interference. For the transverse Ising model it is possible to simulate quenches which end deeply in the disordered phase. However, attempts to go beyond few-steps Trotterrization lead to the strong increase of the total error, which is mainly due to the errors of two-qubit gates as well as the decoherence processes. Our results thus elucidate limitations of current technology in the modeling of unitary evolution of spin models. Nevertheless, we believe that further technological improvements as well as scaling towards computers with many qubits will indeed allow to realize a modeling of the dynamics difficult to study with conventional supercomputers. We also discuss certain heuristic tricks, which can be used to extract some useful information from the imperfect and "noisy" results of the modeling.

Note that the unitary dynamics of a spin chain has recently been tackled in Ref. \cite{Martinisspin} using 9-qubit quantum processor within the algorithmic modeling. However, Ref. \cite{Martinisspin} was focused on adiabatic evolution in connection with the quantum annealing, which was originally introduced for analog quantum machines \cite{analog1,analog2,analog3}. It was argued in Ref. \cite{Martinisspin} that quantum annealing is more attractive to realize digitally, i.e., using programmable quantum computers based on the standard gate model. In contrast, in the present paper, we concentrate on far-from-equilibrium nonadiabatic dynamics based on the same model.

Our paper is organized as follows. In Section II we describe the architecture of 5-qubit IBMqx4 chip and discuss physical systems most suitable for quantum computation with this chip, which belong to the family of the central spin models. We show how degrees of freedom of the modeled system can be mapped on degrees of freedom of the physical device and how various initial conditions, which include both entangled and disentangled states, can be encoded into the device taking into account limitations in its topology. We then perform modeling of unitary evolution, discuss errors as well as the dependence of the dynamics on the initial state. In Section III, we consider the implementation of the unitary dynamics of the transverse-field Ising model in 16-qubit IBMqx5 chip. Both chain and ladder configurations are considered. Section IV summarizes our results and conclusions.

\section{Central spin model and 5-qubit quantum computer}

The algorithmic simulation of spin dynamics consists of several steps. As a first step, one has to encode the initial conditions to the chip. For state-of-art processors, the results can be improved by taking into account a difference in errors of quantum gates associated with particular physical qubits and by choosing the optimal set of physical qubits of the device to perform the simulation, which also implies that the topology of the device has to be accounted for. At the second stage, the dynamics is implemented using Trotter expansion of the evolution operator. At the third state, measurements of qubits are performed. The whole algorithm is repeated many times, which is $8102$ in our case for each point. Let us first utilize this strategy using 5-qubit processor IBMqx4.

\begin{figure}[h]
\center
    \includegraphics[width=0.45\linewidth]{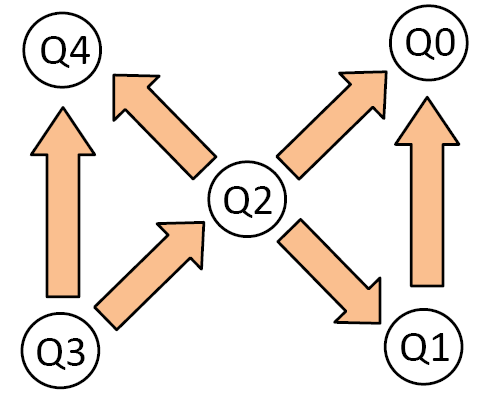}
    \caption{
        \label{chip}
        (Color online) Schematic view of IBMqx4 chip. Two-qubit gates and their directions are shown by arrows (see in the text).}
\end{figure}

\subsection{Mapping to the quantum chip}

The structure of the superconducting quantum processor IBMqx4 is shown schematically in Fig. 1. The central qubit Q2 is connected by controlled-NOT gates (CNOTs) to four remaining qubits Q0, Q1, Q3, and Q4, as indicated in Fig. 1 by arrows. Each arrow points from the control qubit to the target qubit and its directions is caused by the details of the engineering process; in quantum algorithms CNOTs can be easily inverted using additional Hadamard gates.

The most evident approach is to associate this topology with a central spin model, which describes an ensemble of spin-1/2 particles, the central particle interacting with all other particles (bath), see, e.g., Ref. \cite{Prokofiev}. The one-to-one correspondence between the topology of the chip and of the central model allows to minimize the number of quantum gates of the algorithm and thus to reduce errors of our modeling. The interaction between the central particle and particles of the bath can be implemented in a digital way using CNOTs, which connect central qubit with four others (for details on implementation of interaction, see below). The particles of the bath either do not interact with each other directly or do interact: additional CNOTs between Q3 and Q4 as well as between Q0 and Q1 can be used to implement digitally pairwise interaction of corresponding particles, whose quantum states are encoded into these four qubits. CNOTs between any two qubits can be also used to construct entangled quantum states of these two qubits. Thus, entangled initial states of the modeled system can be directly mapped to the entangled states of physical qubits. This scheme suggests one-to-one correspondence between the states of modeled system of particles and physical qubits of the device, as dictated by the interaction term of the modeled Hamiltonian as well as the structure of the initial state of the modeled system. However, in practice, we use several different schemes. Particularly, in order to model the dynamics of the system with three entangled particles, we have to go beyond simple one-to-one mapping due to certain limitation of real chip topology.


The simplest relevant model from the class of central spin models is XX central spin model with the Hamiltonian of the form
\begin{eqnarray}
H_{cs}=  \sum_{j=1}^L  \epsilon_j (\sigma_{j,z}+1/2)+ \epsilon_c (\sigma_{c,z}+1/2) + \nonumber \\
g \sum_{j=1}^L (\sigma_{c}^+ \sigma_{j}^- + \sigma_{c}^- \sigma_{j}^+),
\label{Hamiltoniancentralspin}
\end{eqnarray}
where $\sigma_{j,z}$ and $\sigma_{j}^\pm$ are Pauli operators associated with particles of the bath, while $\sigma_{c,z}$ and $\sigma_{c}^\pm$ refer to the central spin; $\epsilon_j$ and $\epsilon_c$ are excitation energies of spins of the bath, which do not interact with each other directly, whereas $g$ is the interaction constant between the central spin and each spin of the bath. This Hamiltonian is integrable, which facilitates the analysis of our results. For simplicity, we hereafter assume that all excitation energies are the same and switch to the rotating frame. The Hamiltonian we model reads as
\begin{eqnarray}
H =  g \sum_{j=1}^L (\sigma_{c}^+ \sigma_{j}^- + \sigma_{c}^- \sigma_{j}^+).
\label{Hamiltoniancentralspin1}
\end{eqnarray}
In order to bring it to the form conventional in the field of quantum computation we rewrite operators $\sigma^+$ and $\sigma^-$ through operators $\sigma^{x}$ and $\sigma^{y}$. After simple algebra, we ultimately represent (\ref{Hamiltoniancentralspin1}) in an equivalent form as
\begin{eqnarray}
H = \frac{g}{2} \sum_{j=1}^L (\sigma_{c}^{x} \sigma_{j}^{x} + \sigma_{c}^{y} \sigma_{j}^{y}).
\label{Hamiltonian}
\end{eqnarray}

We are going to address three different realizations of spin systems each being characterized by its own initial condition. Let us first consider the system of two particles of the bath coupled to the central particle. We assume that the initial state of the whole system is an unexcited central spin and entangled particles of the bath. This state is given by
\begin{eqnarray}
\Psi_I (0) = |{\downarrow}\rangle \otimes \frac{1}{\sqrt{2}} \left(|{\downarrow} {\uparrow}\rangle + e^{i\varphi}|{\uparrow} {\downarrow}\rangle\right),
\label{init1}
\end{eqnarray}
which is parameterized by a single tunable parameter $\varphi$. We further refer the two-particle entangled state appearing in Eq. (\ref{init1}) to as 2PES. The free evolution of the system must be highly sensitive to the phase parameter $\varphi$. Particularly, at $\varphi=\pi$ the excitation has to be blocked in the environment without being transferred to the central spin. The excitation blockade is caused by quantum interference effect leading to the mutual cancellation of contributions from two different qubits: $H\Psi_I (0)=g |{\uparrow}\rangle \otimes \frac{1}{\sqrt{2}} \left(|{\downarrow} {\downarrow}\rangle + e^{i\varphi}|{\downarrow} {\downarrow}\rangle\right)
= g |{\uparrow}\rangle \otimes \frac{1}{\sqrt{2}} |{\downarrow} {\downarrow}\rangle \left(1 + e^{i\varphi}\right)= 0$ at $\varphi=\pi$. This means that the evolution operator in the rotating frame does not change $\Psi_I (0)$ at $\varphi=\pi$. In contrast, Rabi-like oscillations, i.e.,  free excitation transfer between the subsystem of periphery spins and the central spin do occur at $\varphi \neq \pi$.

\begin{figure}[h]
\center
    \includegraphics[width=0.5\linewidth]{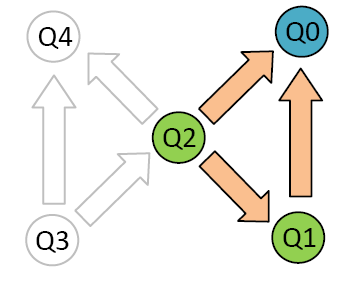}
    \caption{ (Color online) Mapping between the modeled system of two spin-1/2 particles coupled to the central spin and elements of the physical device. Green circles denote physical qubits Q1 and Q2 which encode quantum states of the two particles, while blue circle denote physical qubit Q0 used to encode central spin state. Qubits and CNOTs, which are not used in the algorithm, are shown in grey.
        \label{algo1}
    }
\end{figure}

\begin{figure}[h]
\center
    \includegraphics[width=0.5\linewidth]{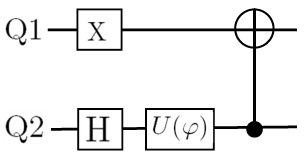}
    \caption{
        \label{2AES}
        Quantum circuit for preparation of the two-particle entangled state 2PES (\ref{init1}).}
\end{figure}

We associate the state of two particles with the quantum states of two physical qubits via one-to-one correspondence, as shown in Fig. \ref{algo1}. Q1 and Q2 are used to encode states of particles of the bath, whereas Q0 encodes quantum states of the central spin. CNOT between Q0 and Q1 as well as between Q0 and Q2 will be used to implement interaction between the central spin and the bath, as described below. CNOT between Q1 and Q2 is used to create an entangled state 2PES of these two qubits. Note that multiple choices to perform mapping between the three spin-1/2 particles and physical qubits do exist for IBMqx4 chip even under restrictions originating from the form of interaction term of Hamiltonian (\ref{Hamiltonian}) and available CNOTs of the chip. We used the optimized one, which is based on minimization of total error induced by CNOT gates in our experiments, since different CNOTs of the chip exhibit different errors. The initial entangled state of two qubits can be prepared using the circuit shown in Fig. \ref{2AES}, where $U(\varphi)=\begin{bmatrix}
1 & 0 \\
0 & e^{i\varphi}%
\end{bmatrix}$: H and X are Hadamard and Pauli-X gates, respectively.

\begin{figure}[h]
\center
    \includegraphics[width=0.95\linewidth]{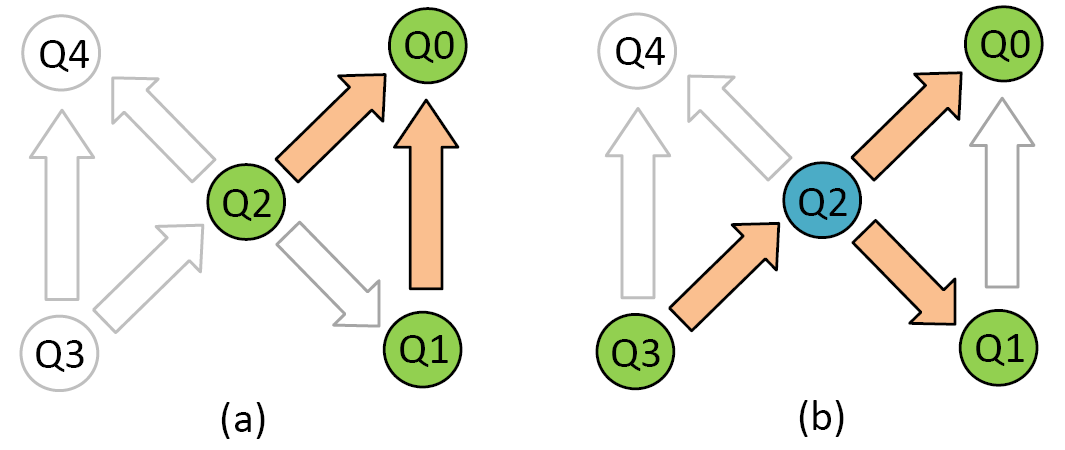}
    \caption{
        \label{algo2}
        (Color online) Mapping between the modeled system of three spin-1/2 particles coupled to the central spin and elements of the physical device. Green circles denote physical qubits which encode quantum states of three particles, while blue circle denote physical qubit used to encode central spin state. At the first stage (a) an entangled state of three physical qubits is created, which encode the entangled state of three particles via one-to-one correspondence. At the second stage (b) the state of the central qubit Q2 is transferred to the unused qubit Q3, while Q2 is further used to encode quantum state of the central spin and to simulate the dynamics. Qubits and CNOTs, which are not used in a given stage, are shown in grey.}
\end{figure}

\begin{figure}[h]\center
    \includegraphics[width=0.6\linewidth]{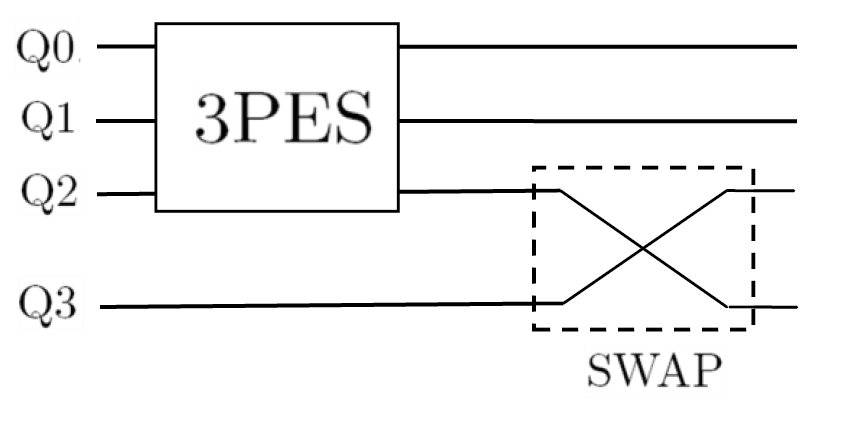}
    \caption{
        \label{3AES}
        Quantum circuit for preparation of the three-particle entangled state 3PES (\ref{init2}) encoded into the physical device.}
\end{figure}

We also consider the system of three spin-1/2 particles coupled to the central spin, the initial state of the system being
\begin{eqnarray}
\Psi_{II} (0) = |{\downarrow}\rangle \otimes \frac{1}{\sqrt{6}}\left(|{\downarrow} {\downarrow} {\uparrow}\rangle - 2e^{i\chi}|{\downarrow} {\uparrow} {\downarrow} \rangle + | {\uparrow} {\downarrow} {\downarrow}\rangle\right),
\label{init2}
\end{eqnarray}
where $\chi$ is tunable phase parameter. The entangled state of three particles in Eq. (\ref{init2}) is further referred to as 3PES. At $\chi=0$ the excitation must be blocked without being transferred to the central qubit due to the cancelation of contributions of three different qubits, since $H\Psi_{II} (0)=g |{\uparrow}\rangle \otimes \frac{1}{\sqrt{6}} |{\downarrow} {\downarrow} {\downarrow} \rangle \left(1- 2 e^{i\chi}+1\right)$. The demonstration of the excitation blockade for the three-particle entangled state (\ref{init2}) is more difficult than that for the two-particle entangled state (\ref{init1}), since larger number of physical qubits has to be involved providing mutually canceling contributions. The mapping of $\Psi_{II} (0)$ to the physical system is also less straightforward compared to the case of 2PES due to the limitations of the chip topology. The difficulty is in the fact that in order to create 3PES (\ref{init2}) it is necessary to use three physical qubits with two CNOT gates between them. Therefore, central physical qubit Q2 of the chip has to be utilized. However, the same qubit has to encode quantum state of the central spin, since there must be three CNOT gates between it and three other qubits in order to model an interaction of the central spin and three particles of the bath. For this reason, the initial state (\ref{init2}) is prepared in two steps, as shown in Fig. \ref{algo2}. At the first stage, 3PES is created using Q0, Q1, and Q2 with the help of two CNOTs. At the second stage, the state of Q2 is transferred to Q3 using the standard SWAP two-qubit gate, which can be composed of three CNOTs, while Q2 is utilized to encode the state of the central spin. Thus, Q0, Q1, and Q3 are ultimately used to encode quantum state 3PES of three particles of the bath, whereas Q2 encodes quantum states of the central spin. The whole quantum circuit for preparation of the initial state is shown schematically in Fig. \ref{3AES}. The initial block used to prepare 3PES encoded into qubits Q0, Q1, and Q2 is described in Appendix A.

\begin{figure}[h]\center
    \includegraphics[width=0.45\linewidth]{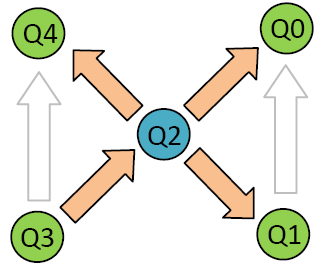}
    \caption{ (Color online) Mapping between the modeled system of four spin-1/2 particles coupled to the central spin and elements of physical device. Green circles denote physical qubits which encode states of four particles, while blue circle denote physical qubit used to encode state of the central spin. Unused CNOTs are shown in grey.
        \label{algo3}
    }
\end{figure}

Next, we address the system of up to four spin-1/2 particles coupled to the central spin, the initial state of the bath being disentangled. We assume that bath initially contains unexcited particles, whereas central spin is excited
\begin{eqnarray}
\Psi_{III} (0) = |{\uparrow} \rangle \otimes |{\downarrow} \ldots {\downarrow} \rangle.
\label{init3}
\end{eqnarray}
Mapping between the modeled system and elements of the chip are shown in Fig. \ref{algo3}. Periphery qubits Q0, Q1, Q3, and Q4 are used to encode states of four particles of the bath, whereas Q2 encodes the state of the central spin.

\begin{figure}[h]\center
    \includegraphics[width=0.95\linewidth]{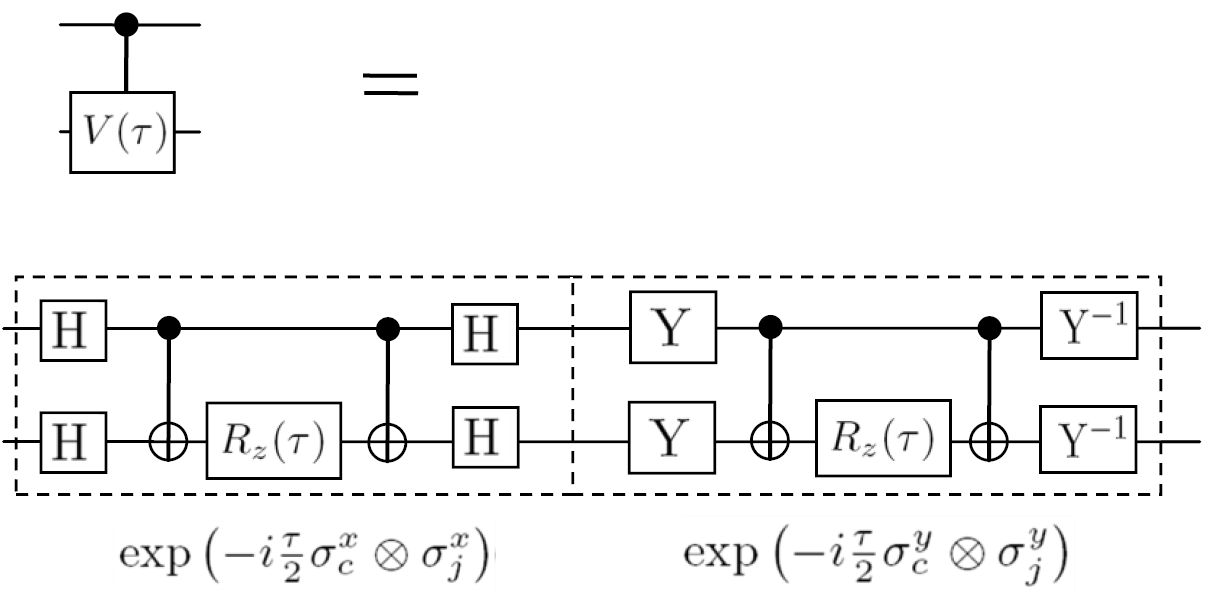}
    \caption{
        \label{interaction}
        Quantum circuit for $\exp\left(-i\frac{\tau}{2} \sigma_{c}^{x} \otimes \sigma_{j}^{x}\right) \exp\left(-i\frac{\tau}{2} \sigma_{c}^{y} \otimes \sigma_{j}^{y}\right)$ (see in the text).}
\end{figure}

\subsection{Trotterization of the evolution operator}

We are now in the position to implement the dynamics of modeled system in a digital way using Trotterization. The free evolution of the system starting from the initial state $| \Psi (0) \rangle$ is given by the standard formula
\begin{eqnarray}
\Psi (t) = e^{-iHt}\Psi (0).
\label{evol}
\end{eqnarray}
Under the expansion involving Trotter number $N$, this expression is rewritten in the approximate form as
\begin{eqnarray}
\Psi (\tau) \approx  \left[ \prod_{j=1}^{L} e^{-i \frac{\tau}{2N} \sigma_{c}^{x}\otimes \sigma_{j}^{x}} e^{-i\frac{\tau}{2N} \sigma_{c}^{y} \otimes \sigma_{j}^{y}} \right]^N \Psi (0),
\label{evoltrot}
\end{eqnarray}
where dimensionless time $\tau=gt$ was introduced. $\sigma_{c}$ and $\sigma_{j}$ now refer to those physical qubits, which encode states of the central spin and particles of the bath, respectively. Eq. (\ref{evoltrot}) is exact in the limit $N\rightarrow\infty$. At $N \sim 1$, it remains accurate provided $\tau \ll 1$ ($t \ll 1/g$), i.e., during the initial period of system's free evolution.

Each gate the form $\exp\left(-i\frac{\tau}{2} \sigma_{c}^{x,y,z} \otimes \sigma_{j}^{x,y,z}\right)$ can be represented in a usual
way thought the CNOT gate entangling two qubits as well as the single-qubit gate
$R_z(\tau)=\exp\left(-i\frac{\tau}{2} \sigma^{z} \right)$ in the appropriate basis.
$R_z(\tau)$ is expressed through the standard IBMqx4 gate $U_3$ and Hadamard gate H as H$ U_3(\theta = \tau, \varphi = - \pi/2, \lambda = \pi/2)$H.

Figure \ref{interaction} presents quantum circuit for $\exp\left(-i\frac{\tau}{2} \sigma_{c}^{x} \otimes \sigma_{j}^{x}\right) \exp\left(-i \frac{\tau}{2} \sigma_{c}^{y} \otimes \sigma_{j}^{y}\right)$, whereas Pauli-Y gate is expressed as $U_3(\theta = - \pi/2, \varphi = - \pi/2, \lambda = \pi/2)$.

Thus, we trace the time evolution of the system starting from three different initial conditions encoded into the real physical device and using the Trotter expansion with different Trotter numbers. In particular, we concentrate on the dynamics of the mean population of the excited state of the central particle by measuring population of qubit $n_c$, which encodes central spin states, at different values of dimensionless time $\tau$. The whole quantum circuits corresponding to three initial conditions are presented in Appendix B (for simplicity, for a single Trotter number, $N=1$).

\begin{figure}[h]\center
    \includegraphics[width=0.48\linewidth]{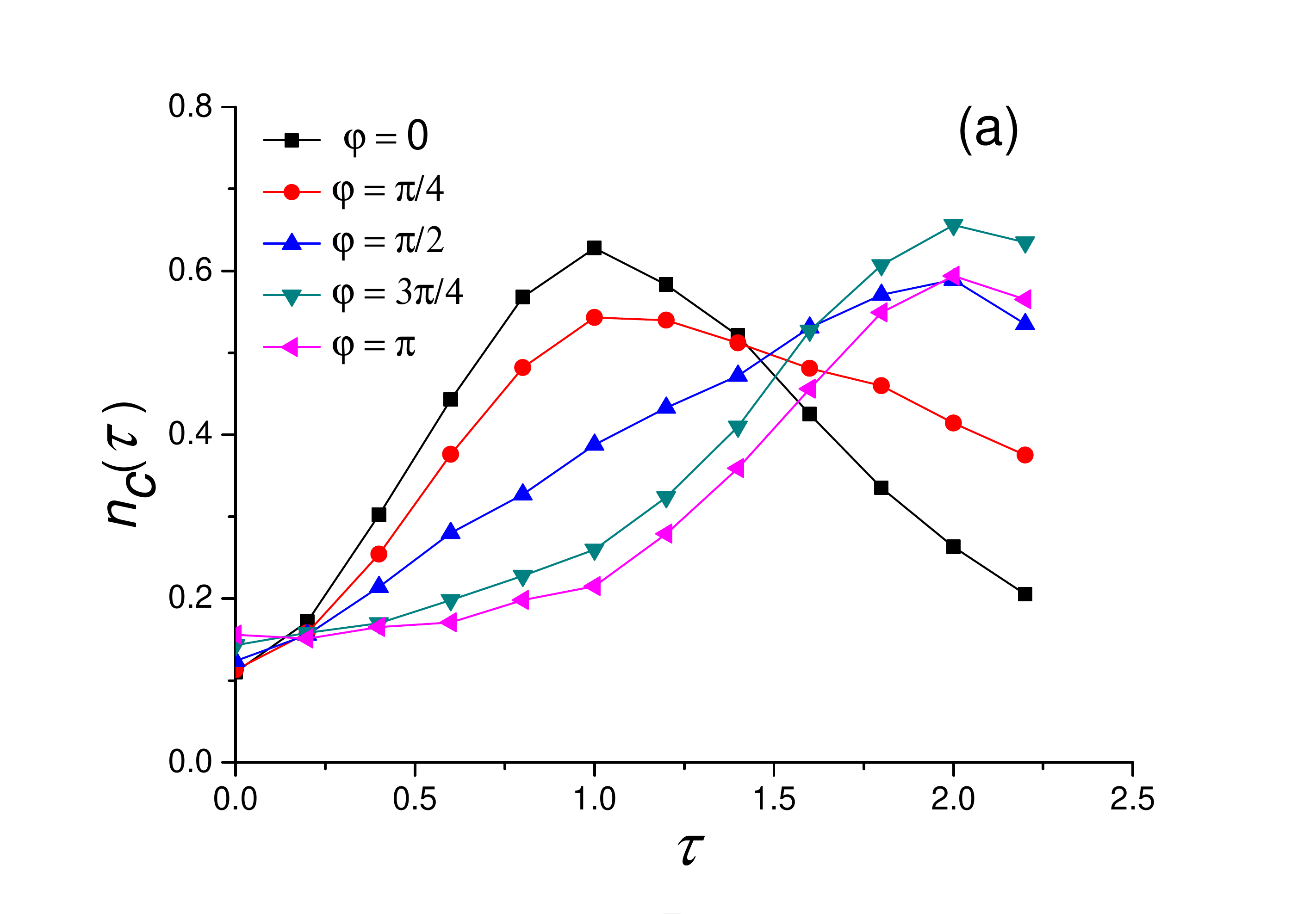}
    \includegraphics[width=0.48\linewidth]{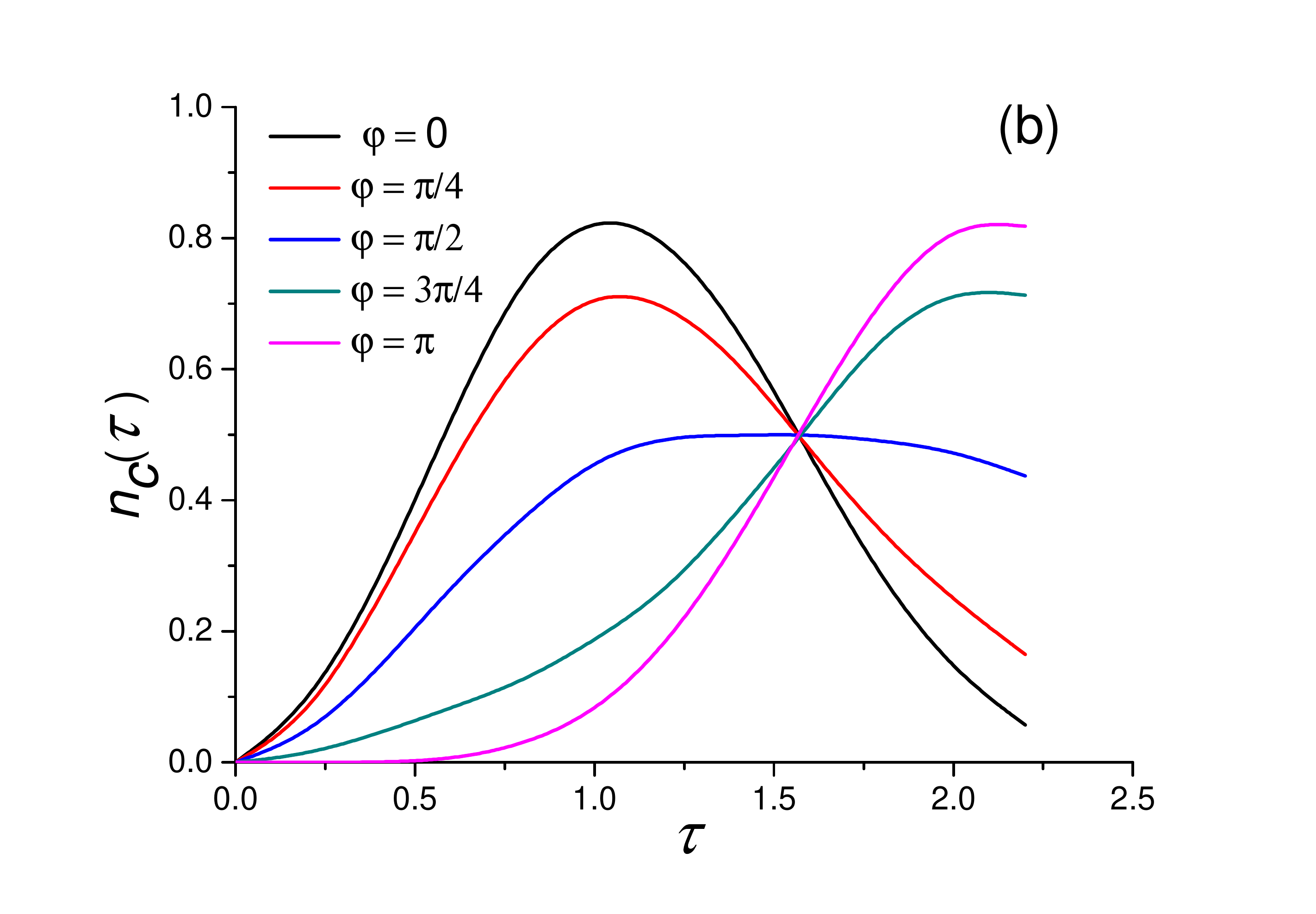}
    \caption{
        \label{results1}
        (Color online) The results of our experiment (a) and theory (b) for the mean population $n_c$ of the excited state of central particle as a function of the dimensionless time $\tau$ for the Trotter number $N=1$. The initial state of the system is two-particle entangled state of the bath and unexcited central spin. Different curves correspond to different values of phase parameter $\varphi$ entering the initial state (\ref{init1}). }
\end{figure}

\begin{figure}\center
    \includegraphics[width=0.48\linewidth]{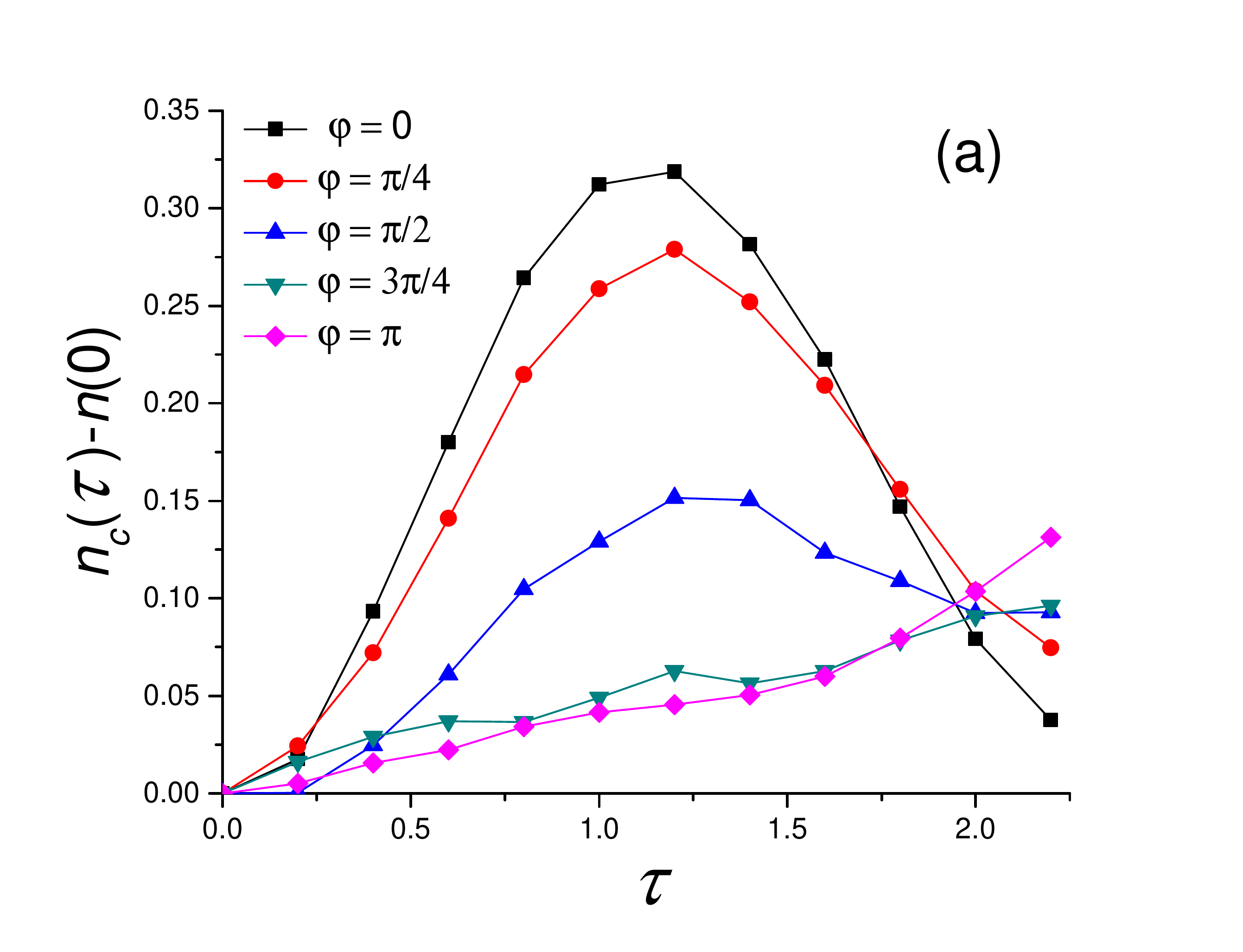}
    \includegraphics[width=0.48\linewidth]{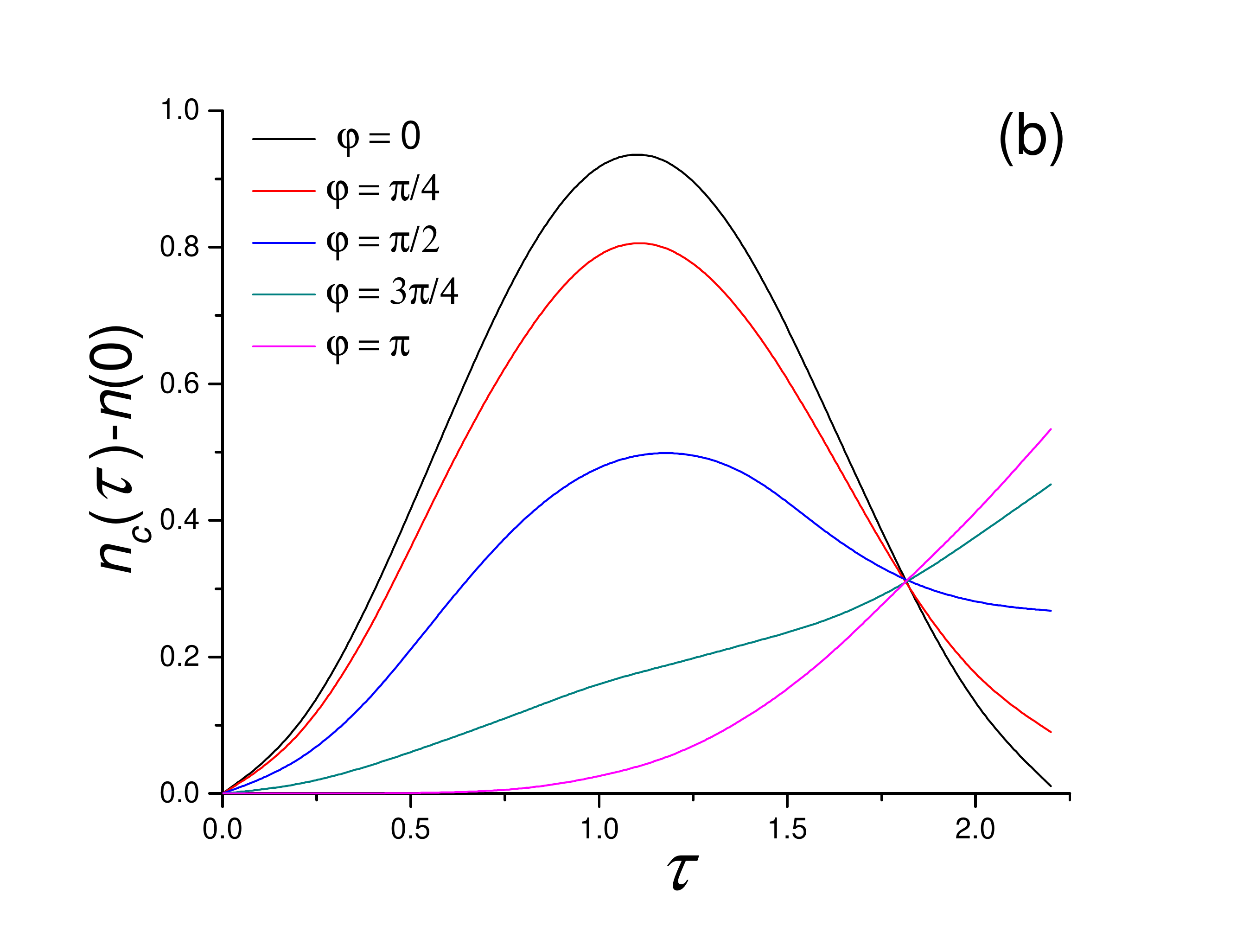}
    \caption{
        \label{results2trotter}
        (Color online) The results of our experiment (a) and theory (b) for $\Delta n_c (\tau)$ as a function of the dimensionless time $\tau$ for the Trotter number $N=2$. Different curves correspond to different values of phase parameter $\varphi$ entering the initial state (\ref{init1}). }
\end{figure}

\subsection{Comparison between the experiment and theory}

Now we compare the experimental and theoretical results for different initial states introduced above and Trotter numbers $N$. Let us stress that such a comparison is always made by us for approximations of the same level, i.e., for the same $N$ in Eq. (\ref{evoltrot}) used both for the experiment and for the theory. The errors of the physical device grow with the increase of $N$ due to the increasing length of the algorithm, which makes the effects of quantum gates imperfections and decoherence more and more significant. However, Trotterization errors associated with finite $N$ in Eq. (\ref{evoltrot}) decrease as $N$ grows.

Figure \ref{results1} shows the evolution of the mean population of the excited state of the central particle starting from the initial state (\ref{init1}) for several values of the phase parameter $\varphi$ and under the one-step Trotter decomposition, $N=1$. Figure \ref{results1} (a) corresponds to the experimental results obtained with IBMqx4 quantum computer, while Fig. \ref{results1} (b) shows the results of the theoretical predictions based on the same approximation -- one-step Trotter decomposition of the evolution operator. Both results should be identical in the case of an ideal quantum computer (no decoherence as well as gate and readout errors). The theoretical results can be readily found explicitly using Eq. (\ref{evoltrot}) or they can be obtained from IBM simulator (classical computer) available in the IBM Q online system, which performs the same calculations numerically and can be used for the analysis of experimental results. We here work with data from this simulator because of its usability.

The sensitivity of the population of the central spin to the phase parameter $\varphi$ can be understood by adopting quantum optics picture and replacing the central spin by the oscillator, which is exact in the single-excitation regime. Indeed, the value of phase parameter $\varphi$ equal to $\pi$ corresponds to the so called non-radiant or dark state. This nonradiant state is an eigenstate of the Hamiltonian and therefore no mean photon occupation is induced upon the free evolution starting from this state (for the exact solution, i.e., infinite Trotter number). The dynamics seen in Fig. \ref{results1} (b) for $\varphi = \pi$ at relatively large $\tau \sim 1$ is an artifact of one-step Trotter decomposition. This approximation is reliable at $\tau \ll 1$, but it also provides adequate qualitative and even semi-quantitative results at $\tau \lesssim 1$. Tuning $\varphi$ from $\pi$ results in the increased coupling of the initial state to the light. This behavior is reflected within the approximation based on the one-step Trotter decomposition: the initial growth of the mean photon number as a function of time becomes stronger when tuning $\varphi$ away from $\pi$ and reaches maximum at $\varphi=0$, see Fig. \ref{results1} (b). The behavior we expect from the theory (one-step Trotter decomposition) is reproduced in the experiments, as seen in Fig. \ref{results1} (a). The dynamics simulated in the experiment is highly sensitive to the tunable phase parameter $\varphi$; the dependencies on $\varphi$ are the same in Figures \ref{results1} (a) and (b). This sensitivity is unambiguous demonstration for the realization of entangled states and quantum interference effects in the device, which reproduces the effect of the excitation blockade in our digital simulation.

\begin{figure}\center
    \includegraphics[width=0.485\linewidth]{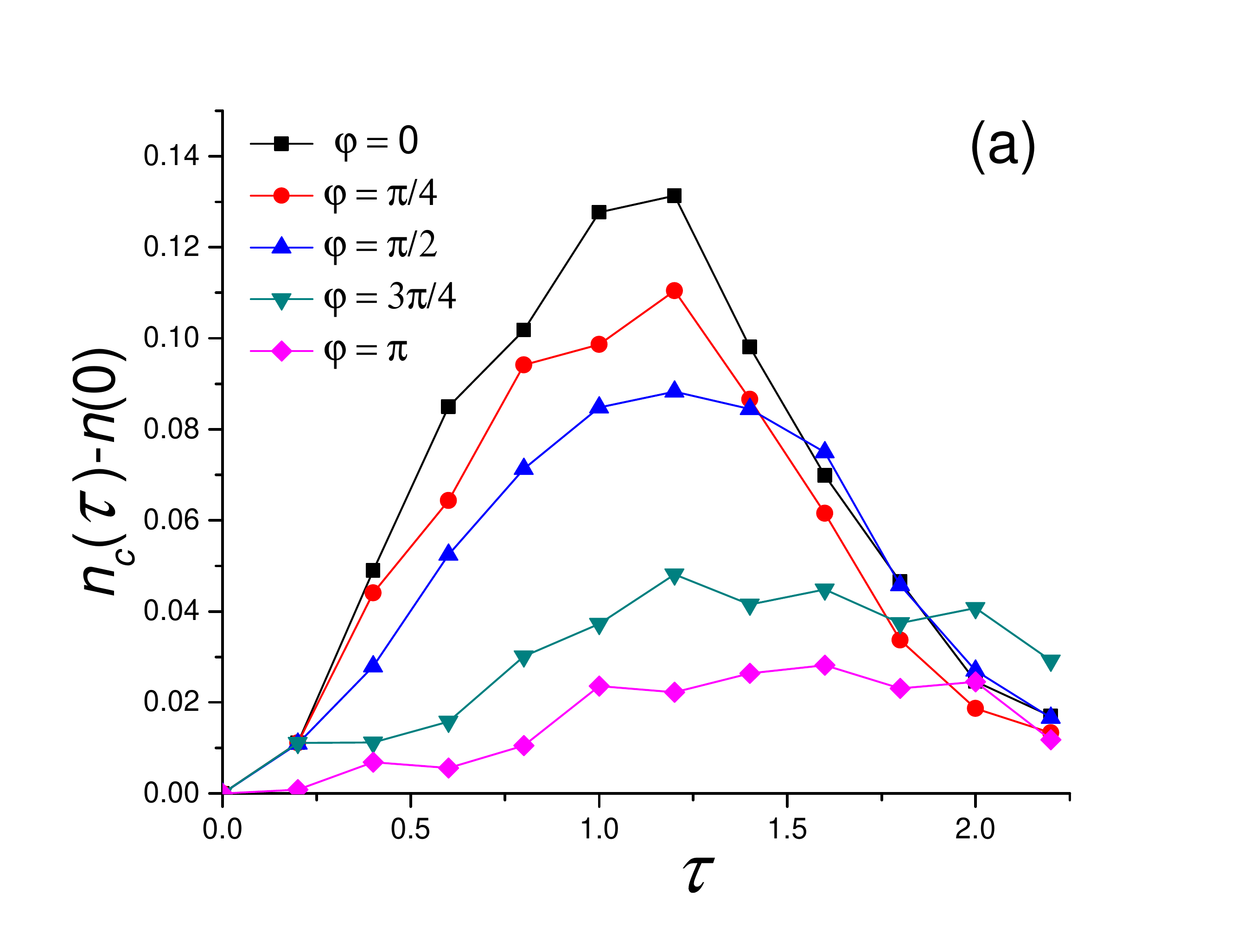}
    \includegraphics[width=0.485\linewidth]{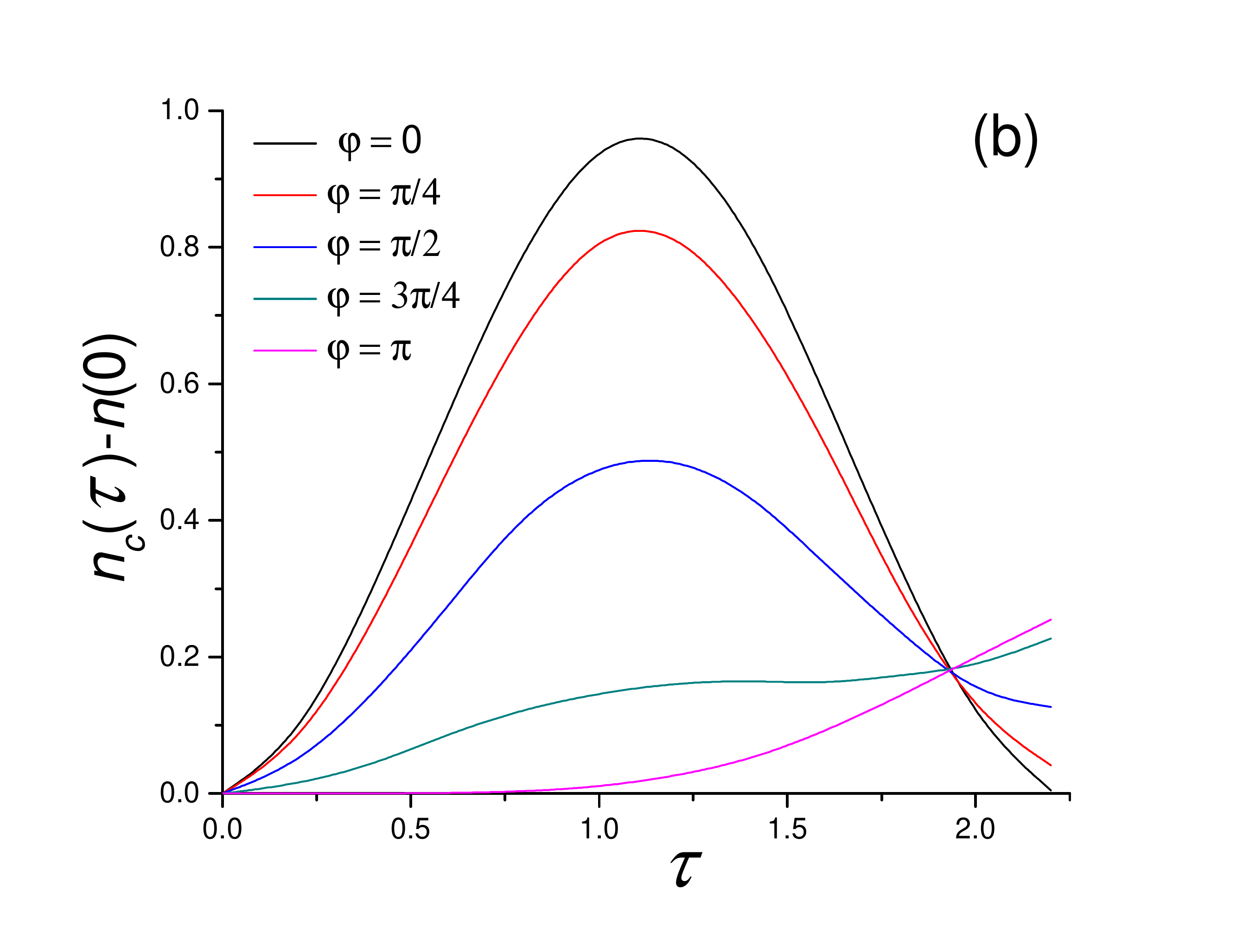}
    \caption{
        \label{results3trotter}
        (Color online) The results of our experiment (a) and theory (b) for $\Delta n_c (\tau)$ as a function of the dimensionless time $\tau$ for the Trotter number $N=3$. Different curves correspond to different values of phase parameter $\varphi$ entering the initial state (\ref{init1}). }
\end{figure}

Nevertheless, there exist significant deviations of the experimental curves compared to the theoretical ones. The reason is mainly in errors of CNOT gates which are typically several percent in IBMqx4 and thus induce quite large error after nearly ten CNOTs per physical qubit. Decoherence of physical qubits also gives noticeable contribution in the time scale of a single run of the algorithm. Of course, one-step Trotterization is also not exact, so it is accurate only at $\tau \ll 1$.

Let us now try to increase Trotter number having in mind a reduction of Trotterization errors in the dynamics. In order to somehow get rid of the nonzero $n_c (\tau=0)$, which appears due to the errors of the physical device, we heuristically suggest to address the difference $\Delta n_c (\tau) = n_c (\tau)-n_c (\tau=0)$. Note that $n_c (\tau=0)$, of course, grows as $N$ is increased. The experimental and theoretical results for $N=2$ and $N=3$ are presented in Figs. \ref{results2trotter} and \ref{results3trotter}, respectively. There is still a qualitative agreement between the theory and the experiment in the sense that the character of the dependencies both on $\tau$ and $\varphi$ are reproduced correctly. However, the qualitative agreement does not exist. In particular, the maximum of $\Delta n_c (\tau)$ as a function of time is nearly 3 and 8  times smaller than the correct values at $N=2$ and 3, respectively. Notice that the theoretical curves for $\varphi=\pi$ become closer and closer to the $x$-axis as $N$ increases, which reflects the reduction of Trotterization errors and more proper realization of the excitation blockade in the Trotterized dynamics.

As a whole, our results demonstrate that despite of large errors, experimental data for several Trotter numbers still contain a lot of valuable information on relative values, so that focusing on differences (perhaps, properly normalized, see the next Section for the illustration) might also has a sense for modeling using "noisy" quantum computers. By using this trick we somehow get rid of the background signal nearly independent on $\tau$.

\begin{figure}\center
    \includegraphics[width=0.485\linewidth]{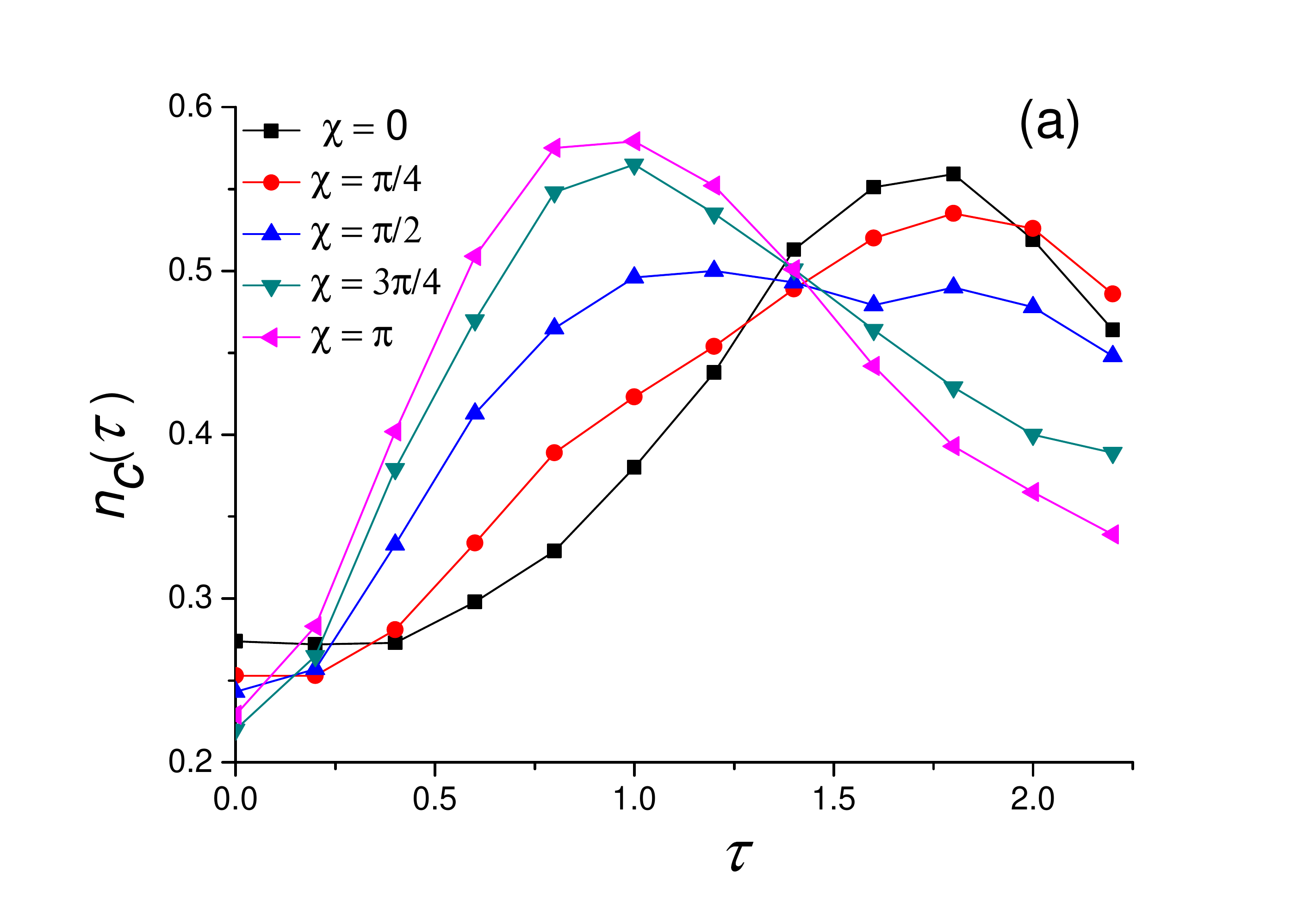}
    \includegraphics[width=0.485\linewidth]{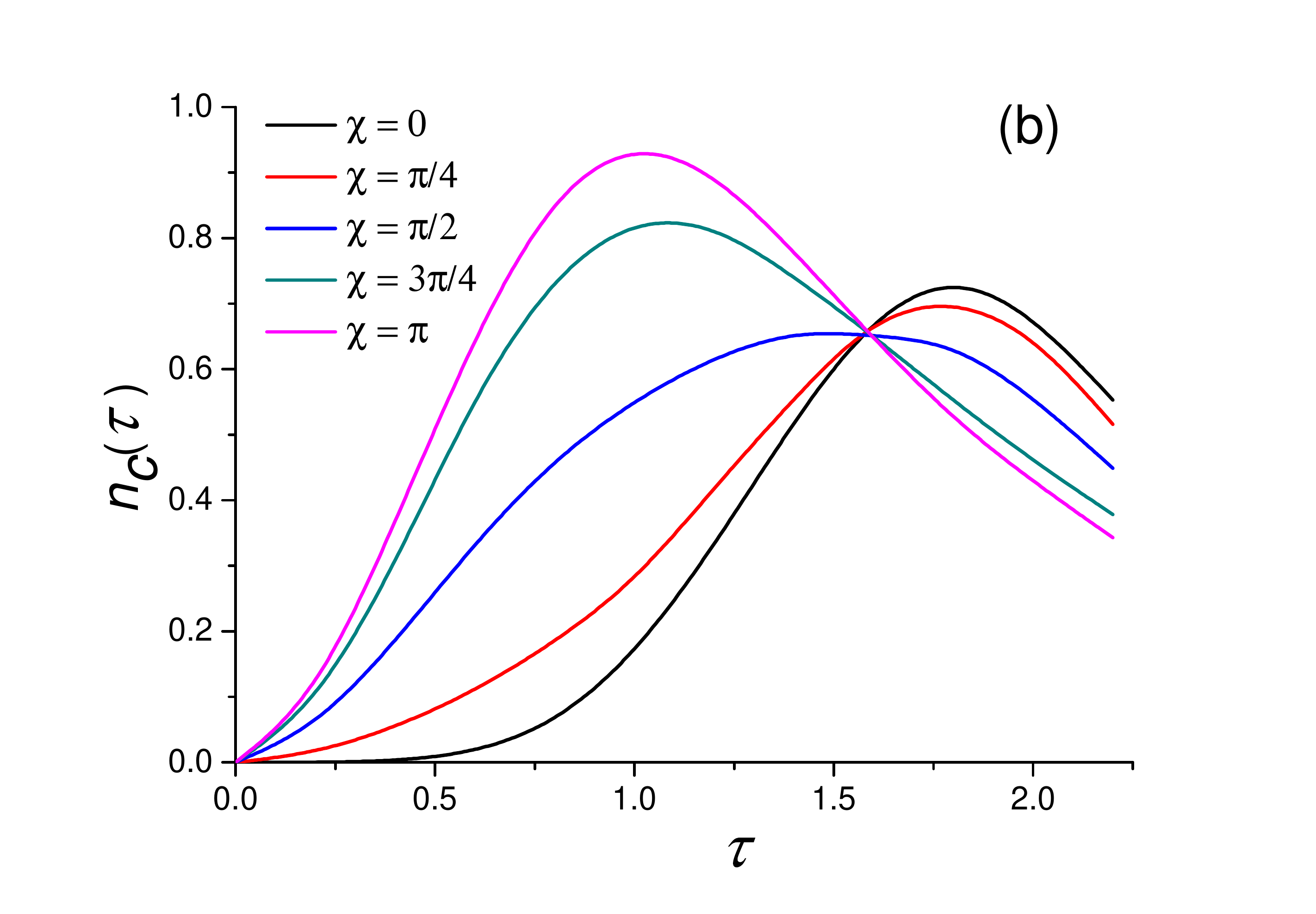}
    \caption{
        \label{results2}
        (Color online) The results of our experiment (a) and theory (b) for the mean population of the excited state of central particle $n_c (\tau)$ as a function of the dimensionless time $\tau$ for the Trotter number $N=1$. The initial state of the system is three-particle entangled state of the bath and unexcited central spin (\ref{init2}). Different curves correspond to different values of phase parameter $\chi$ entering the initial state.  }
\end{figure}

Now we present our results for the initial state (\ref{init2}) involving three entangled particles. Figure \ref{results2} shows the time evolution of the mean population $n_c (\tau)$ of the excited state of central particle for the Trotter number $N=1$. There again exists a good semi-quantitative agreement between the theory and experiment. By theory we again mean an approximation based on one-step Trotter decomposition. However, the sensitivity to the tunable phase parameter $\chi$ is less pronounced for experimental results (a) compared to the theoretical ones (b). The reason is in the increased length of the whole algorithm. Nevertheless, the dynamics is governed by quantum interference effects in this case as well. Particularly, the central spin is weakly occupied for $\chi = 0$ at $\tau$ small, i.e., for the initial three-particle entangled state of the form $|{\downarrow} {\downarrow} {\uparrow}\rangle - 2|{\downarrow} {\uparrow} {\downarrow} \rangle + | {\uparrow} {\downarrow} {\downarrow}\rangle$, which is a quantum superposition of two degenerate dark states $|{\downarrow} {\downarrow} {\uparrow}\rangle - |{\downarrow} {\uparrow} {\downarrow} \rangle$ and $| {\uparrow} {\downarrow} {\downarrow}\rangle - |{\downarrow} {\uparrow} {\downarrow} \rangle$, so that the excitation blockade is again partially reproduced. Due to the increased total errors compared to the first initial condition (\ref{init1}) we studied, we do not implement the algorithm with $N>1$.

\begin{figure}\center
    \includegraphics[width=0.485\linewidth]{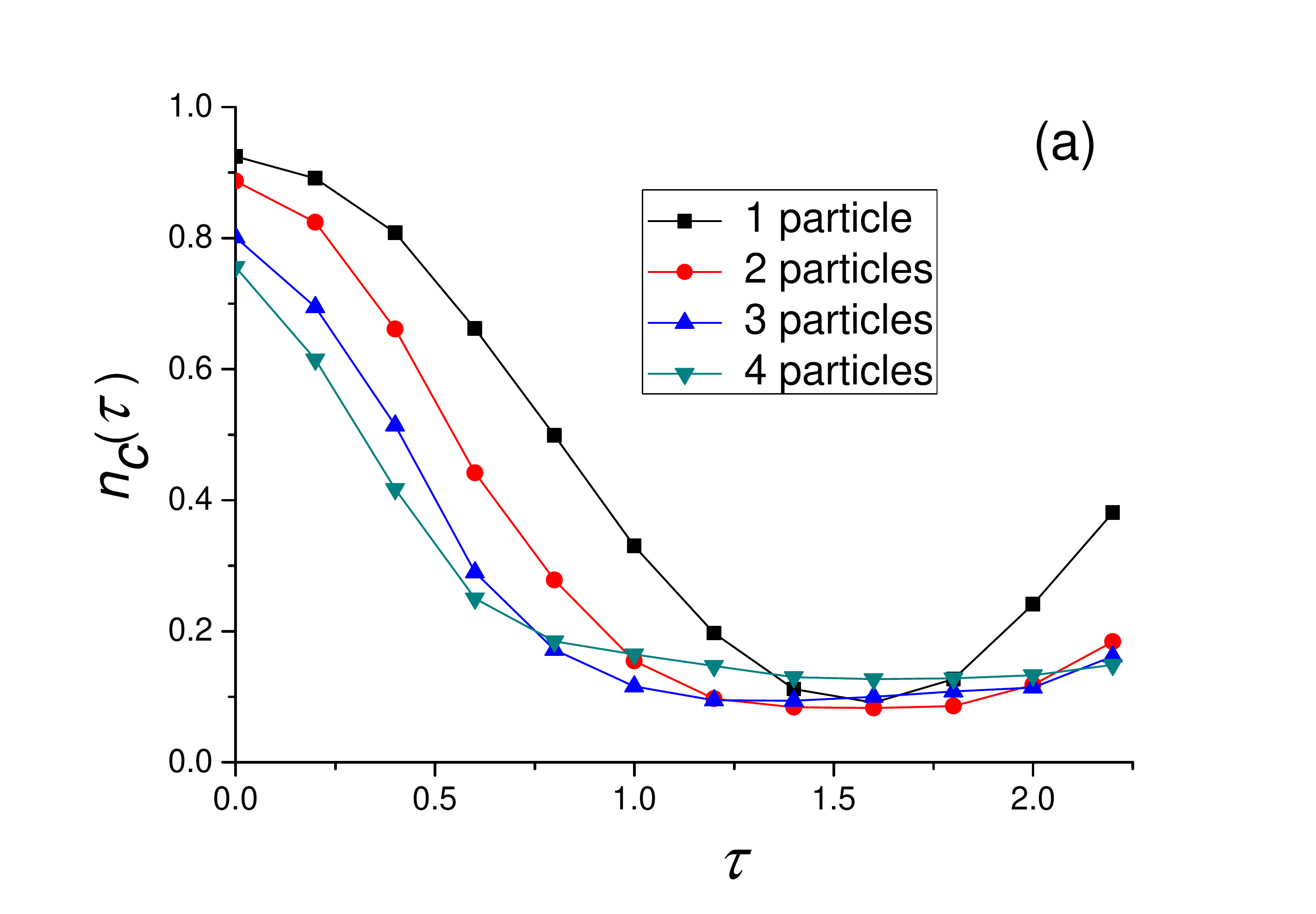}
    \includegraphics[width=0.485\linewidth]{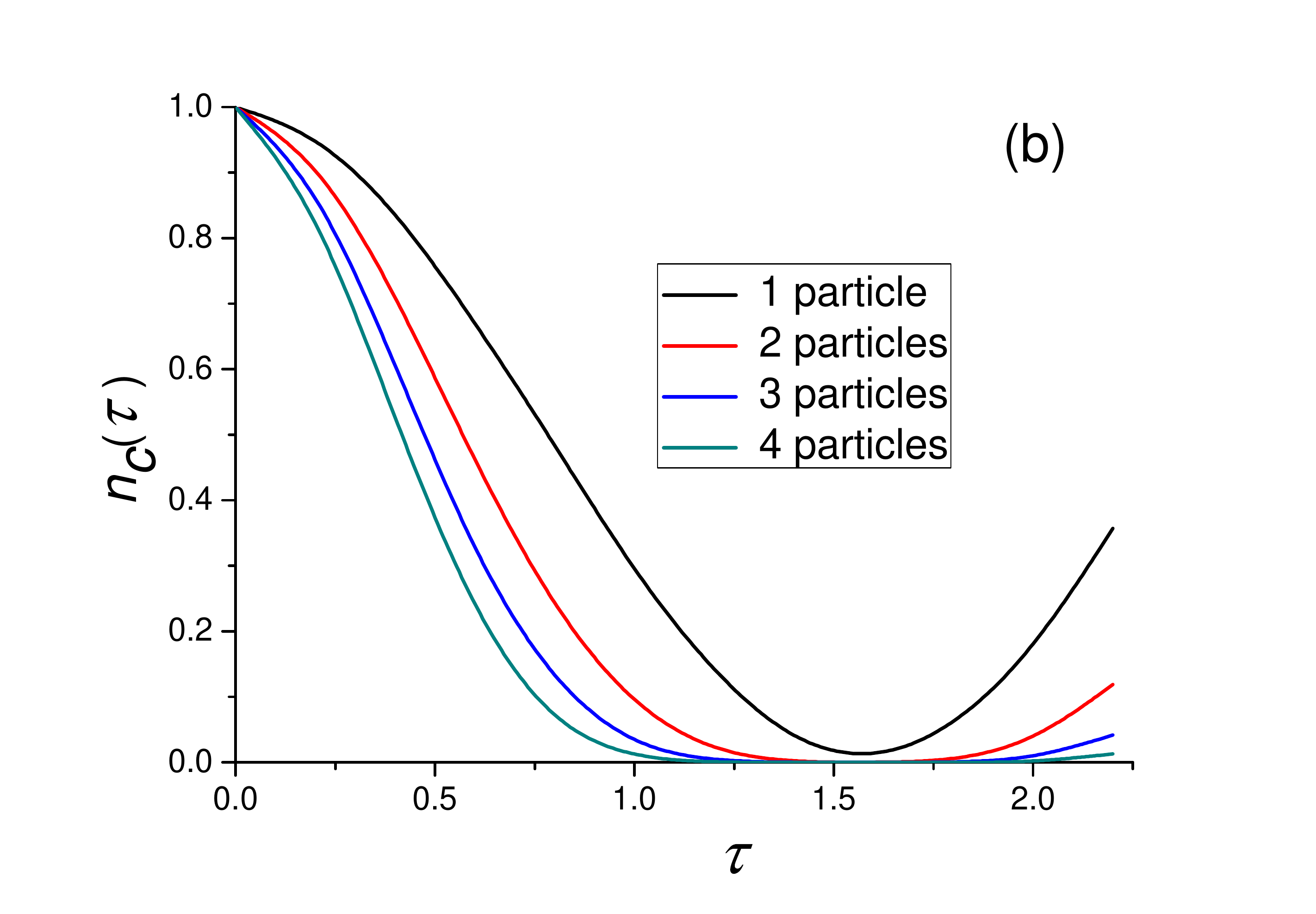}
    \caption{
        \label{results3}
        (Color online) The results of our experiment (a) and theory (b) for the mean population of the excited state of central particle $n_c (\tau)$ as a function of the dimensionless time $\tau$ for the Trotter number $N=1$. The initial state of the system is excited central spin and $L$ unexcited spins of the bath (\ref{init3}). Different curves correspond to different values of $L$. }
\end{figure}

Let us now address dynamics starting from the initially disentangled bath (\ref{init3}): central spin excited and all spins of the bath unexcited. Time evolution of the population of the excited state of central particle for different numbers of spins of the bath, from 1 to 4, is shown in Fig. \ref{results3} using both the experimental (a) and theoretical (b) data at Trotter number $N=1$. Adopting quantum optics understanding, within the exact solution, i.e., infinite number of Trotter steps, one would expect Rabi oscillations between the central spin and the collective spin constructed from individual spins of the bath. Thus, the initial excitation should freely transfer from central spin to the bath and back since no initial entanglement is present in the bath, which could block such a transfer. Moreover, a period of Rabi oscillations is expected to be proportional to $1/\sqrt{L}$, since the interaction energy between the central spin and the bath is enhanced as $g \sqrt{L}$. Fig. \ref{results3} (b) shows that such a collective behavior also exists under the one-step Trotter expansion. Comparing Fig. \ref{results3} (b) with the experimental results (a) obtained by quantum computer using the same one-step Trotter decomposition embedded in the algorithm, we see that cooperative behavior is indeed reproduced in experiments: the first derivative of $n_c(\tau)$ with respect to $\tau$ at $\tau=0$ depends on the number of particles in a correct way being nearly proportional to $\sqrt{L}$, which is a feature typical for collective Rabi oscillations.

Note that the important peculiarity of the central spin model is that the central spin is connected with all remaining spins, so that many CNOTs have to be applied to the particular physical qubit used to encode quantum states of the central spin, which leads to the fast accumulation of errors in this qubit.

\section{Ising model in a transverse field and 16-qubit quantum computer}

\begin{figure}[h]
\center
    \includegraphics[width=0.95\linewidth]{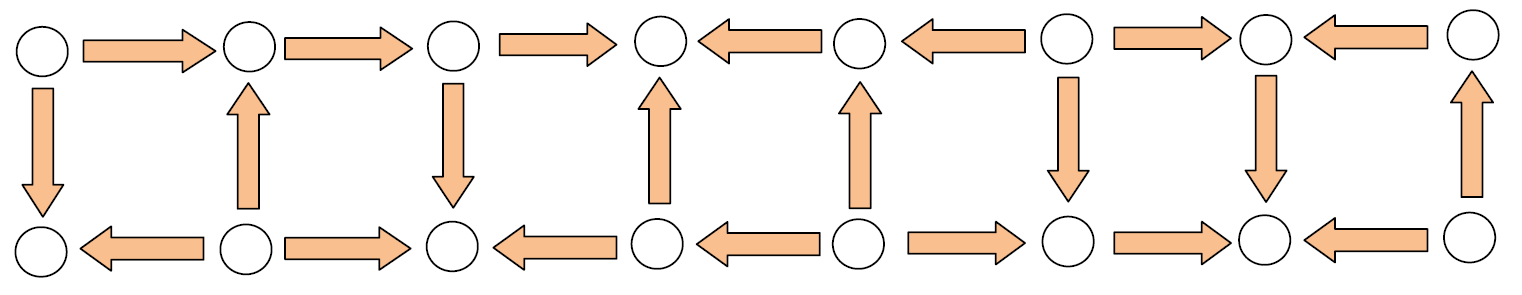}
    \caption{
        \label{chip16}
        (Color online) Schematic view of IBMqx5 chip. Two-qubit gates and their directions are shown by arrows (see in the text).}
\end{figure}

\subsection{Mapping to the quantum chip}

n this Section we discuss an implementation of the dynamics in transverse Ising model using 16-qubit superconducting quantum computer IBMqx5. This model is generally non-integrable beyond the one-dimensional geometry. The structure of the superconducting quantum processor IBMqx5 is shown schematically in Fig. \ref{chip16}. It consists of two rows each containing eight physical qubits. CNOTs can be implemented between any of the two neighboring qubits. This topology is attractive to implement the dynamics of the Heisenberg model, which, in general case, includes all interactions of the $\sigma_{x} \sigma_{x}$, $\sigma_{y} \sigma_{y}$, and $\sigma_{z} \sigma_{z}$ types between nearest neighbors as well as three different interaction constants. Again a the one-to-one correspondence between the topology of the chip and of the Heisenberg chain or ladder allows to minimize the number of quantum gates of the algorithm in our modeling. Physical qubits thus can be used to directly encode spin-1/2 states of the Heisenberg model. While one-dimensional Heisenberg model is integrable, no exact and general solution is known already for ladder-type configurations.

However, the implementation of the Heisenberg model is costly from the viewpoint of the number of CNOTs per each physical qubit and single Trotter step. For "internal" qubits, which have three neighbors, this number is 18 and this is going to lead to the large errors already at $N=1$.

Nevertheless, it is of interest to consider also a family of Ising models, which represent a special case of the Heisenberg model and include only interactions of $\sigma_{z} \sigma_{z}$ type. This is going to require an application of only 6 CNOTs per each "internal" qubit and single Trotter step. The transverse Ising model, as discussed in the Introduction, provides a popular playground to study the far-from-equilibrium dynamics including quenches and explore various fundamental problems of statistical physics such as the phenomena related to the thermalization of closed quantum systems. Individual addressability of qubits allows to create in the chip initial states of a wide class, which include entangled states, while the dependence of thermalization on the initial state is an important issue \cite{Banuls,Gogolin,Polkovnikov1,Polkovnikov2}. Such an opportunity does not exist for analog simulators, which usually are not characterized by so high level of the individual addressability.

\begin{figure}
\center
    \includegraphics[width=0.95\linewidth]{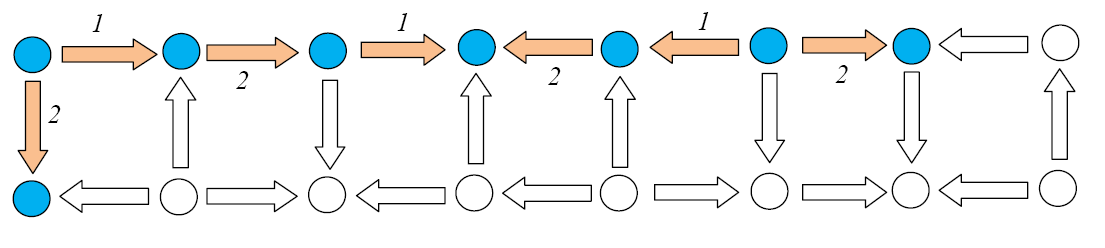}
    \caption{
        \label{scheme8q}
        (Color online)The layout used to simulate 8-spin transverse Ising chain in IBMqx5 chip. Unused qubits and CNOTs are shown in grey. Numbers between two qubits depict parallel operations to implement pairwise interaction between these qubits within each Trotter step.}
\end{figure}

\begin{figure}
\center
    \includegraphics[width=.480\linewidth]{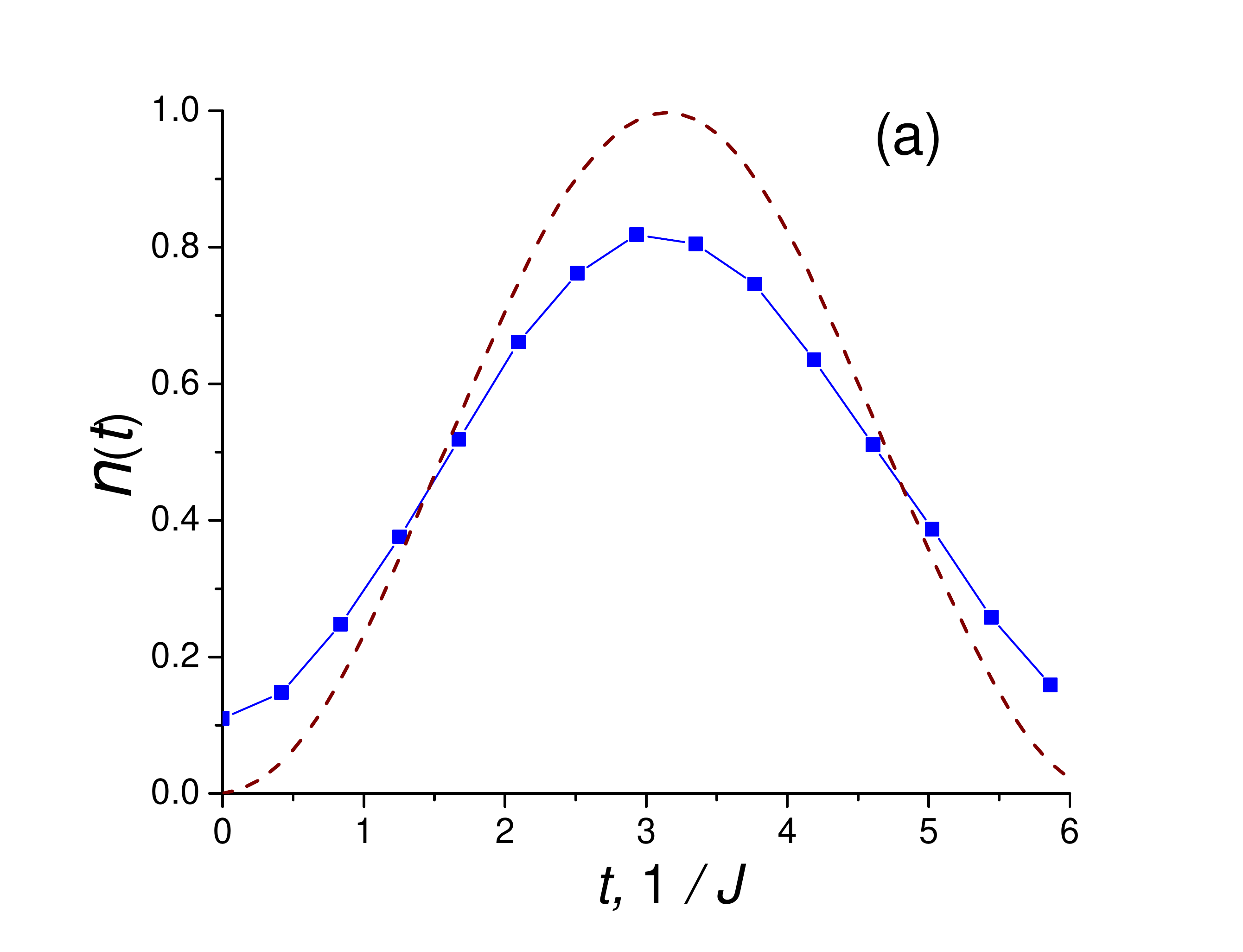}
    \includegraphics[width=.480\linewidth]{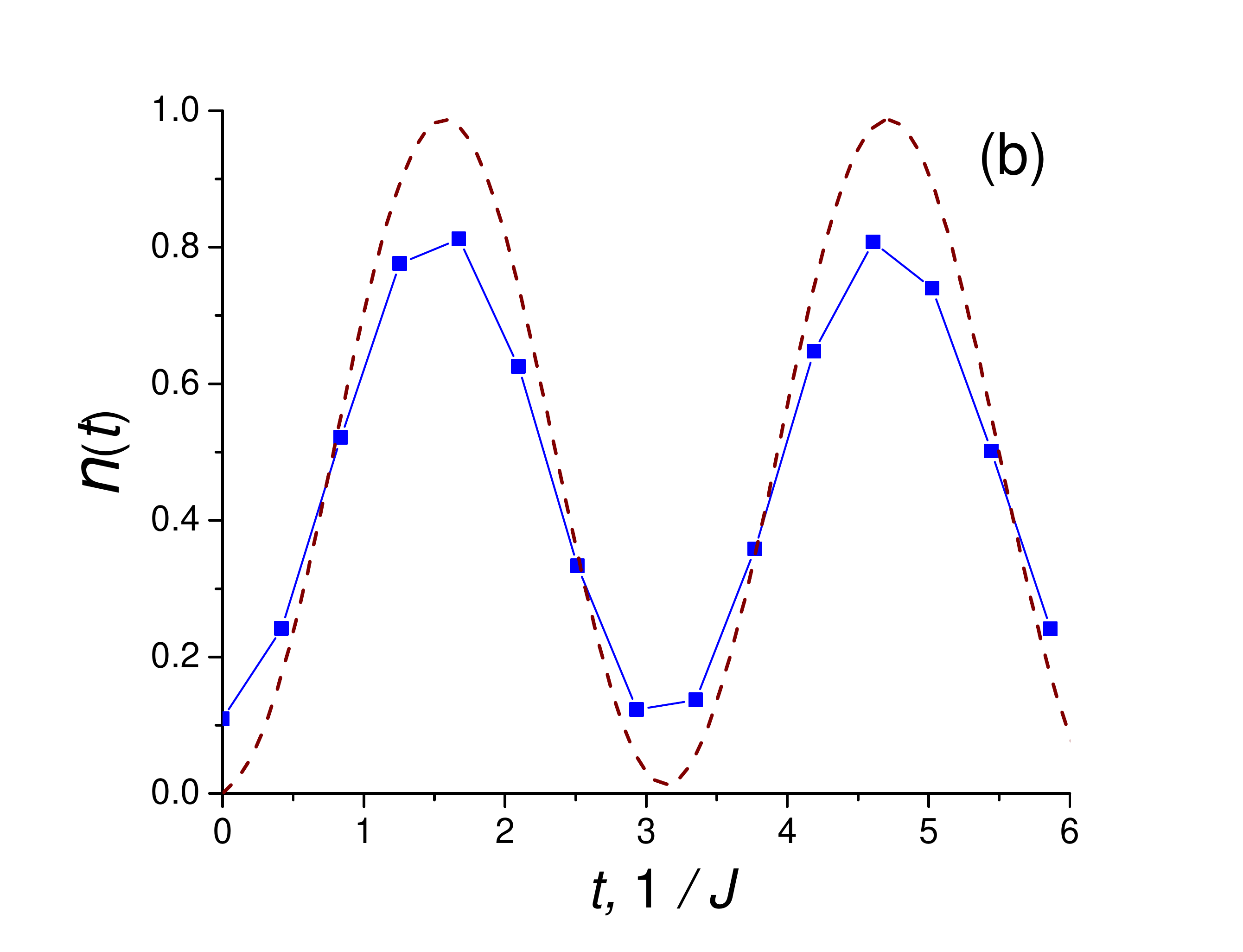}
    \includegraphics[width=.480\linewidth]{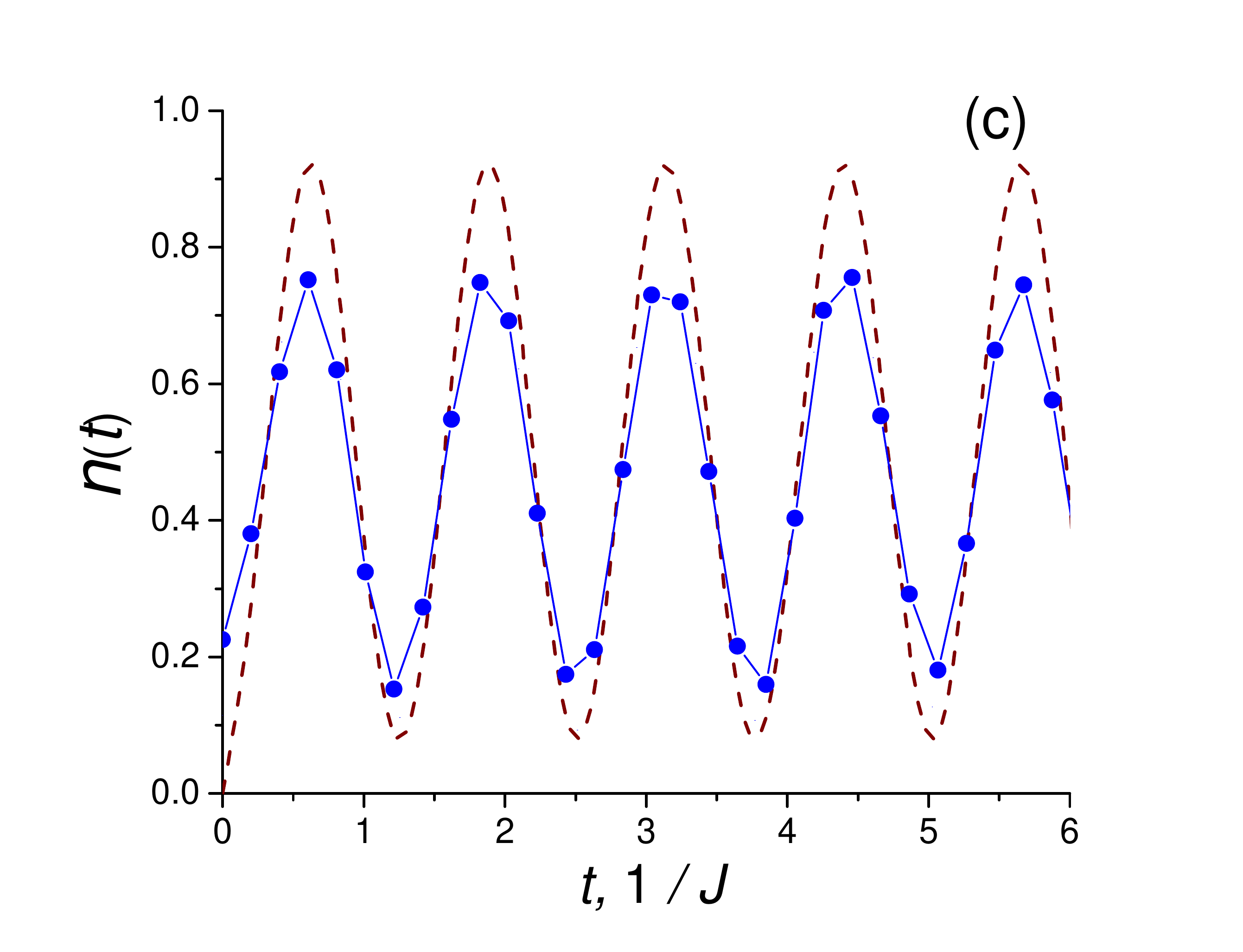}
    \caption{
        \label{Ising8N1}
        (Color online) The results of our experiment (solid blue lines) and theory (dashed brown lines) for the mean occupation $n$ of the upper levels of the 8-spin transverse Ising chain at $\alpha=J$ (a), $\alpha=2J$ (b), $\alpha=5J$ (c) as a function of the time $t$ for the Trotter number $N=1$. }
\end{figure}

\begin{figure}
\center
    \includegraphics[width=.480\linewidth]{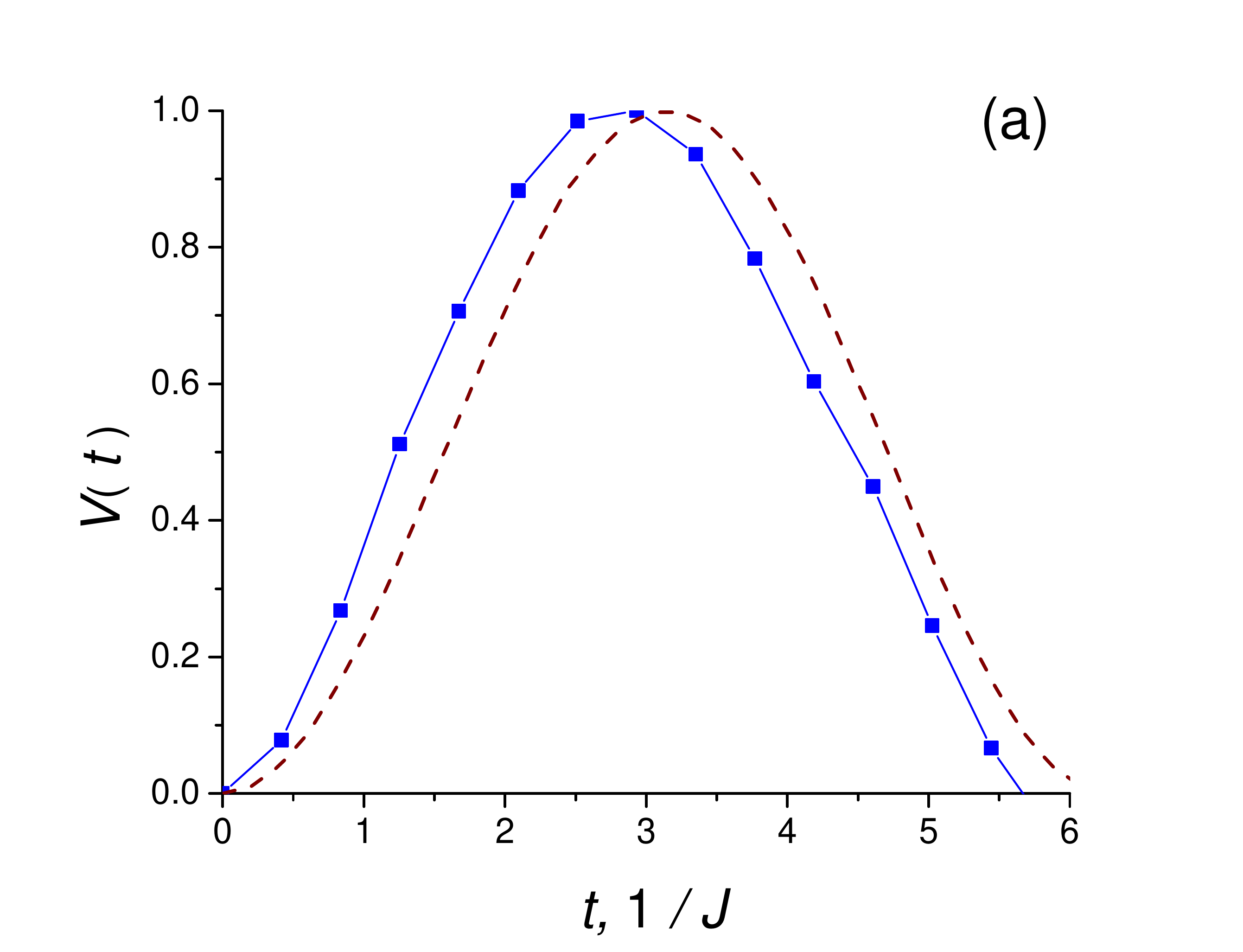}
    \includegraphics[width=.480\linewidth]{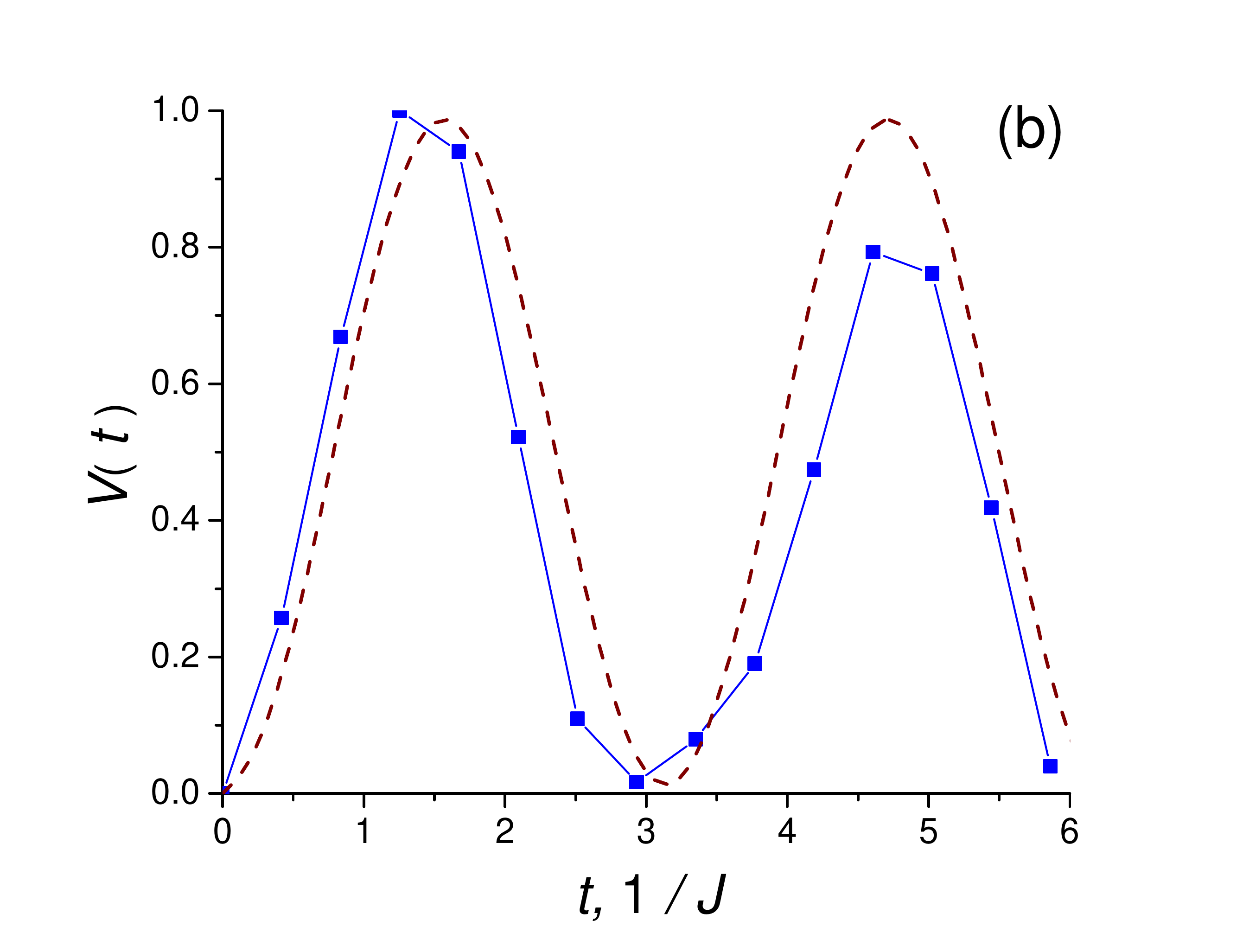}
    \includegraphics[width=.480\linewidth]{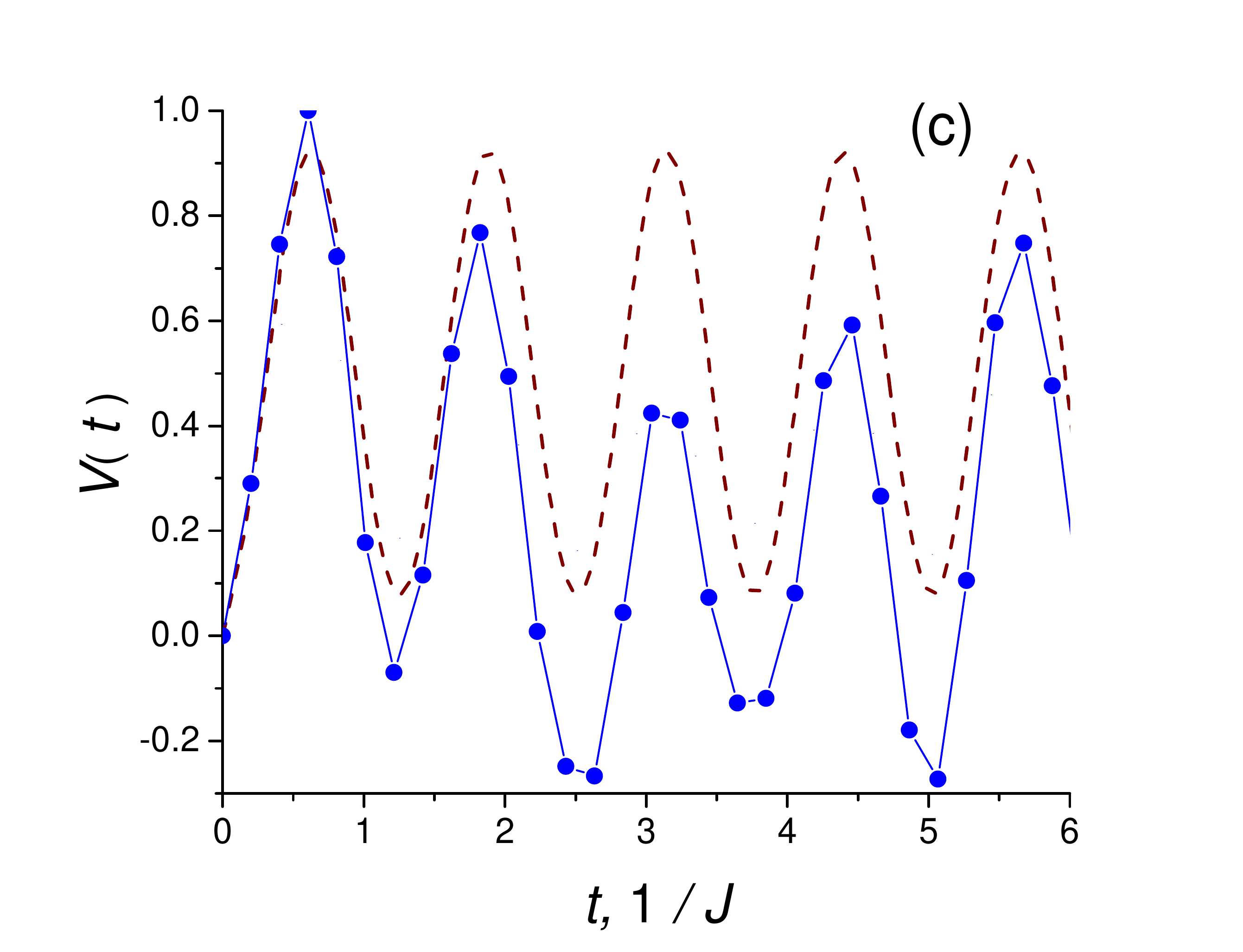}
    \caption{
        \label{Ising8N2}
        (Color online) The results of our experiment (solid blue lines) and theory (dashed brown lines) for $V$ defined in Eq. (\ref{Vtau}) in the case of the 8-spin transverse Ising chain at $\alpha=J$ (a), $\alpha=2J$ (b), $\alpha=5J$ (c) as a function of the time $t$ for the Trotter number $N=2$. }
\end{figure}

The Hamiltonian of the transverse Ising model reads as
\begin{eqnarray}
H = -J\sum_{\langle i,j \rangle}\sigma_{z}^{i}\sigma_{z}^{j}-\alpha \sum_{i}\sigma_{x}^{i},
\label{IsingHam}
\end{eqnarray}
where $\sigma$-operators act on the array of sites (qubits), whereas $\langle i,j \rangle$ refer to the summation over neighboring sites. Note that the transverse Ising model is non-stochastic in the sense that the sign problem appears in Monte-Carlo simulations of this model due to the presence of the second term in the right-hand side of Eq. (\ref{IsingHam}). We also note that, according to the quantum annealing strategy, the parameter $J$ is increased adiabatically, while $\alpha$ is decreased, and the system evolves gradually from the ground state of the model with all qubits in their superposition states to the ground state of the resultant interacting spin model.

In our proof-of-principles experiments, we consider quenches in the transverse Ising model, which occur under the sudden change of system's parameters. We concentrate on the situation, when initially $\alpha=0$, so that the ground state of the system is ferromagnetic, $| {\downarrow} \ldots {\downarrow} \rangle$. Then, $\alpha$ is turned non-adiabatically to some finite value. Time evolution of the system is traced for various final values of $\alpha \geq J$, which is a free parameter in the modeling. In particular, we focus on the mean occupation $n$ of the qubit excited state defined as $n=\langle\sigma_{z}\rangle_q$, where averaging is performed over the qubits (spins) of the system. Note that $\alpha \geq J$ corresponds to quenches ending in the disordered phase of the transverse Ising model, where the dynamics is rather sophisticated \cite{Calabrese}. Two different configurations are considered -- one-dimensional 8-spin chain and 16-spin ladder. In the first case, the number of nearest neighbors for each spin is smaller than the same mean number for the ladder, which results in the reduction in the number of CNOTs per physical qubit per single Trotter step. Therefore, errors due to CNOTs are expected to be smaller for this one-dimensional geometry.

\subsection{Simulation of the spin chain dynamics}

In Fig. \ref{scheme8q}, we show the layout used to model the dynamics of 8-spin transverse-field Ising chain. The unused physical qubits and CNOTs are shown in grey. The "shift" of the first qubit of the chain to the second row of the chip is motivated by the lower readout error in this particular qubit. Each Trotter step includes several stages. Initially, we encode a free part of the Hamiltonian by applying the gate $\exp\left(-i t \sigma^{x}\right)$ on each physical qubit. Next, we encode the term originating from the spin-spin interaction of the Hamiltonian. At the first stage, qubits with "1" between them are entangled simultaneously (in parallel) through $\exp\left(-i\frac{t J}{\alpha} \sigma^{z} \otimes \sigma^{z}\right)$. At the second stage, qubits with "2" are also entangled in parallel through the same operation. This parallel strategy is utilized in order to decrease the total time needed to perform the whole algorithm in order to reduce errors.
Note that IBMqx5 processor is suitable for the modeling of Ising chain of up to 16 spins length, since it is possible to represent such a chain by the unclosed "ring" of physical qubits.

\begin{figure}[h]
\center
    \includegraphics[width=0.95\linewidth]{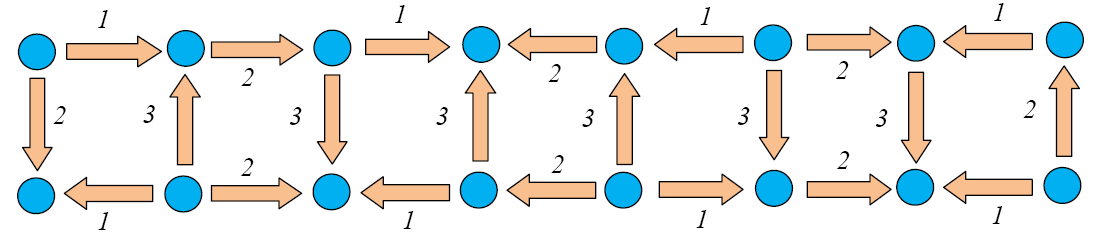}
    \caption{
        \label{scheme16q}
        (Color online) The layout used to simulate 16-spin transverse Ising ladder. Numbers between two qubits depict parallel operations to implement pairwise interaction between these qubits within each Trotter step.}
\end{figure}

Our results for the 8-spin chain are presented in Fig. \ref{Ising8N1} for the Trotter number $N=1$. This figure also contains theoretical results for the approximation of the same level, $N=1$. The inspection of Fig. \ref{Ising8N1} shows that quantum device produces the results, which are semi-quantitatively accurate for $N=1$ and for different values of final $\alpha$ including the ones, which correspond to the strongly disordered phase. Relatively good agreement between the theory and the experiment is explained by the simplicity of the Hamiltonian ($zz$ interaction only) and small number of nearest neighbors for each spin of the model.

However, already for $N=2$ the accuracy drastically reduces. Apparently, the reason is in the increased number of CNOTs applied to each physical qubit. The correct behavior of $n$ is nevertheless still well reproduced on the qualitative level, in similarity to the results for the central spin model described in the preceding Section. We therefore again apply our heuristic approach and instead of focusing on absolute values, we analyze relative quantities. Moreover, we normalize them and ultimately address the quantity defined as
\begin{eqnarray}
V(\tau) = \frac{n(\tau)-n(0)}{\max n(\tau)-n(0)}.
\label{Vtau}
\end{eqnarray}
Both experimental and theoretical results for this quantity are presented in Fig. \ref{Ising8N2} for the Trotter number $N=2$. It shows that there is a good agreement for $V(\tau)$. The raw data for $n(\tau)$ are given in Appendix C. Notice that instead of normalizing by $\max n(\tau)-n(0)$ in the right-hand side of Eq. (\ref{Vtau}), it is possible to use a smoother quantity $\overline{n(\tau)-n(0)}$, where averaging is performed over $\tau$.

The improved quality of the results for $V(\tau)$ and related quantities as compared to raw data can be linked to the fact that erroneous applications of a particular quantum gate of the algorithm leads to the wrong output dependence on $\tau$. These wrong outputs are all different for errors occurring at different quantum gates of the algorithm. After averaging over all such erroneous events, a randomization is expected, i.e., a flat dependence on $\tau$ emerges, provided the number of quantum gates of the algorithm is large. However, there are also runs of algorithms which contain no errors and therefore they produce a correct output against the background. The latter is then eliminated by using Eq. (\ref{Vtau}).

\subsection{Simulation of the spin ladder dynamics}

Figure \ref{scheme16q} provides the layout used to model the dynamics of 16-spin transverse-field Ising ladder. In order to reduce the total time of the algorithm, spin-spin interaction is modeled within each Trotter step at three stages used to entangle qubits simultaneously through  $\exp\left(-i\frac{t J}{\alpha} \sigma^{z} \otimes \sigma^{z}\right)$, as shown by numbers in Fig. \ref{scheme16q}. The results for $V(\tau)$ in the 16-spin ladder are presented in Fig. \ref{Ising16N1}, while the raw data for the dynamics of $n$
are presented in Appendix D. Due to the increased number of nearest neighbors and
CNOTs per physical qubit, errors are high already for the Trotter number $N=1$.
In addition, the whole algorithm for the ladder configuration and a given Trotter number is longer because of the necessity to entangle larger number of qubits. The time of a single run of the algorithm even for a single Trotter step becomes of the order of the mean time $T_2$ of the physical qubits of the chip, so that decoherence processes start to give noticeable contribution to the total error.


In our simulations, the total error per physical qubit can be estimated as $2p_{CNOT}N_{neig}N \nu$, where $p_{CNOT}$ is the CNOT error, $N_{neig}$ is the number of spins participating in the interaction with the given spin, and $\nu$ is a number, which characterizes the complexity of the spin-spin interaction ($\nu$ ranges from 1 for Ising models, which include only $zz$ interaction, to 3 for Heisenberg model, which includes interactions of three types, $xx$, $yy$, and $zz$).

\begin{figure}
\center
    \includegraphics[width=.48\linewidth]{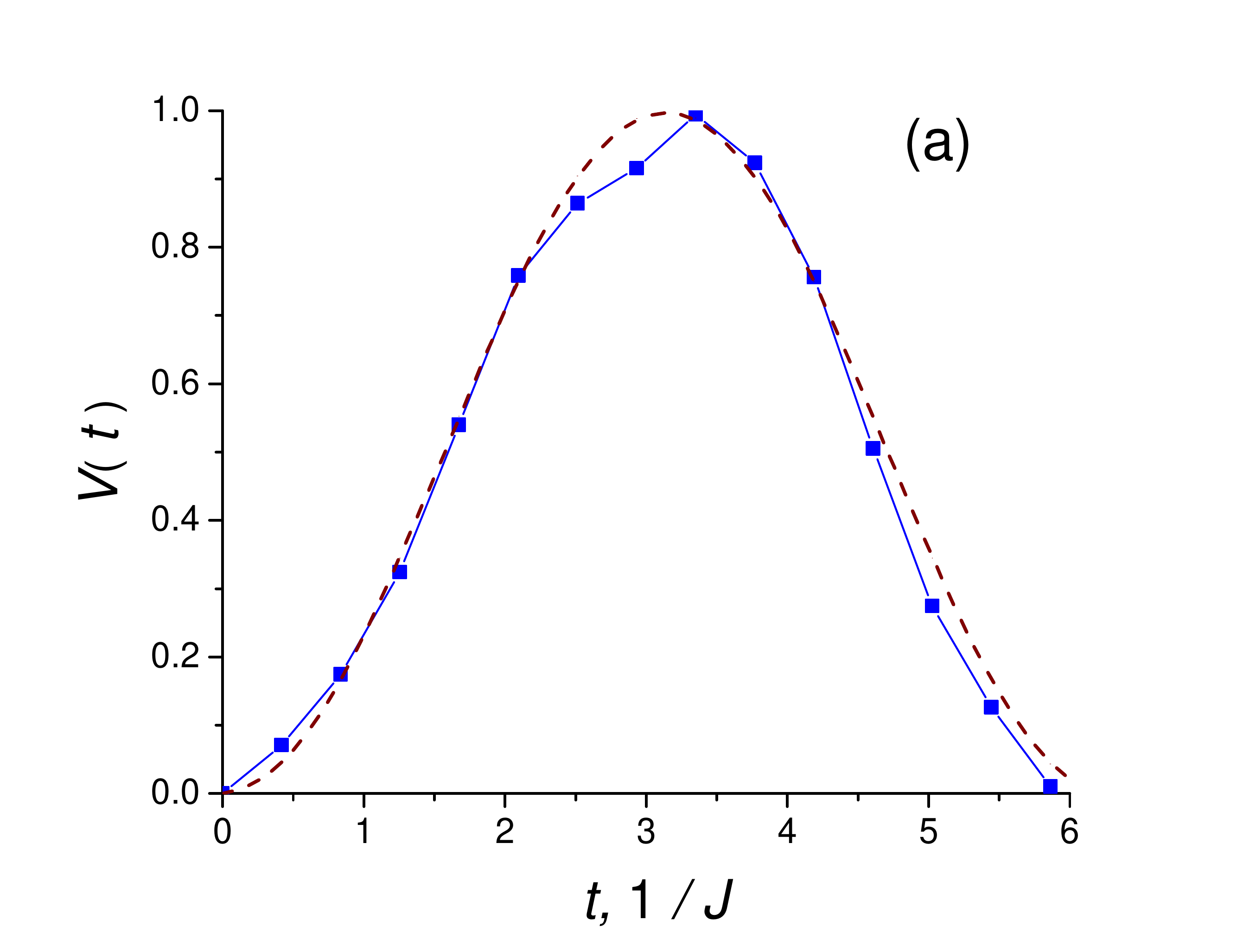}
    \includegraphics[width=.48\linewidth]{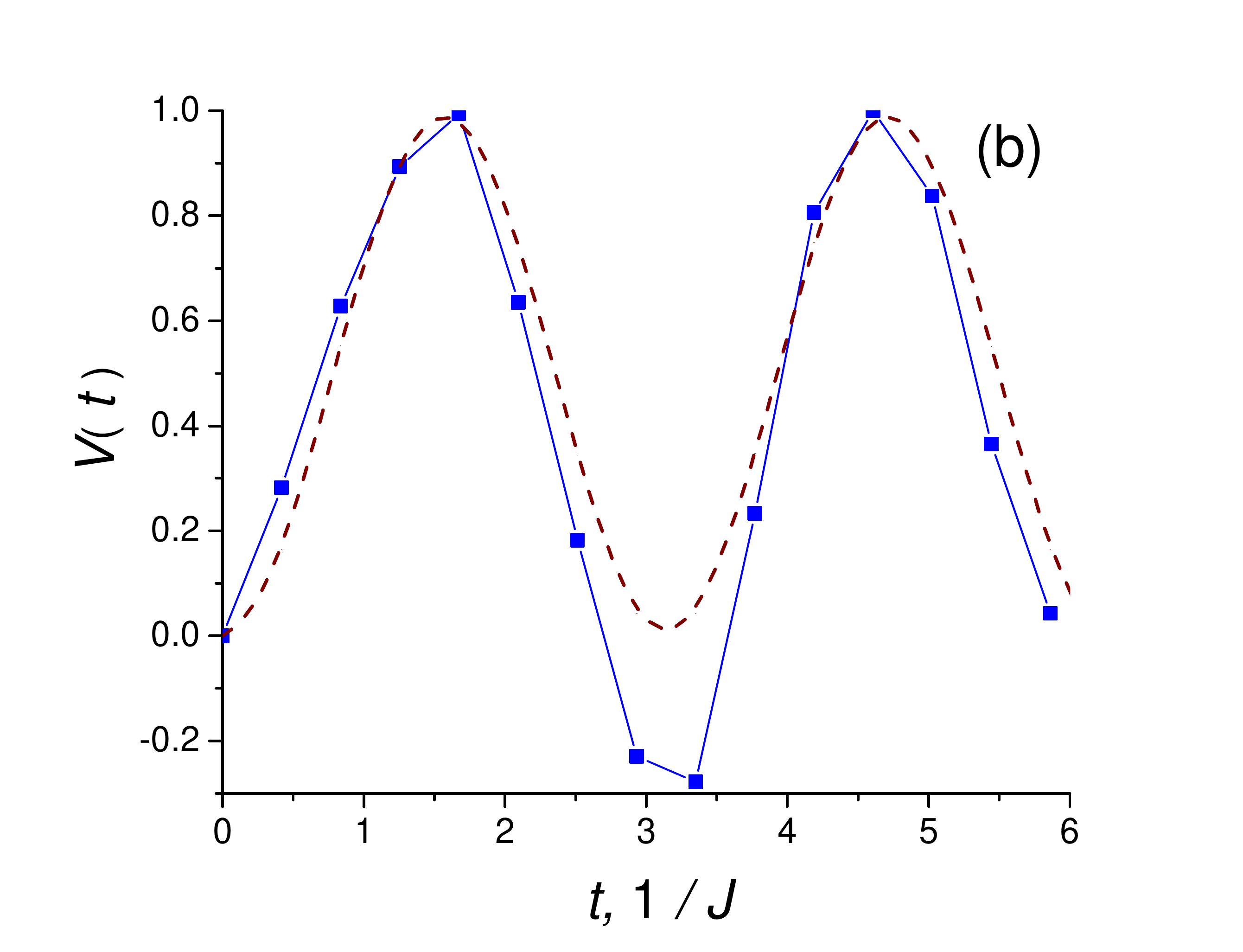}
    \includegraphics[width=.48\linewidth]{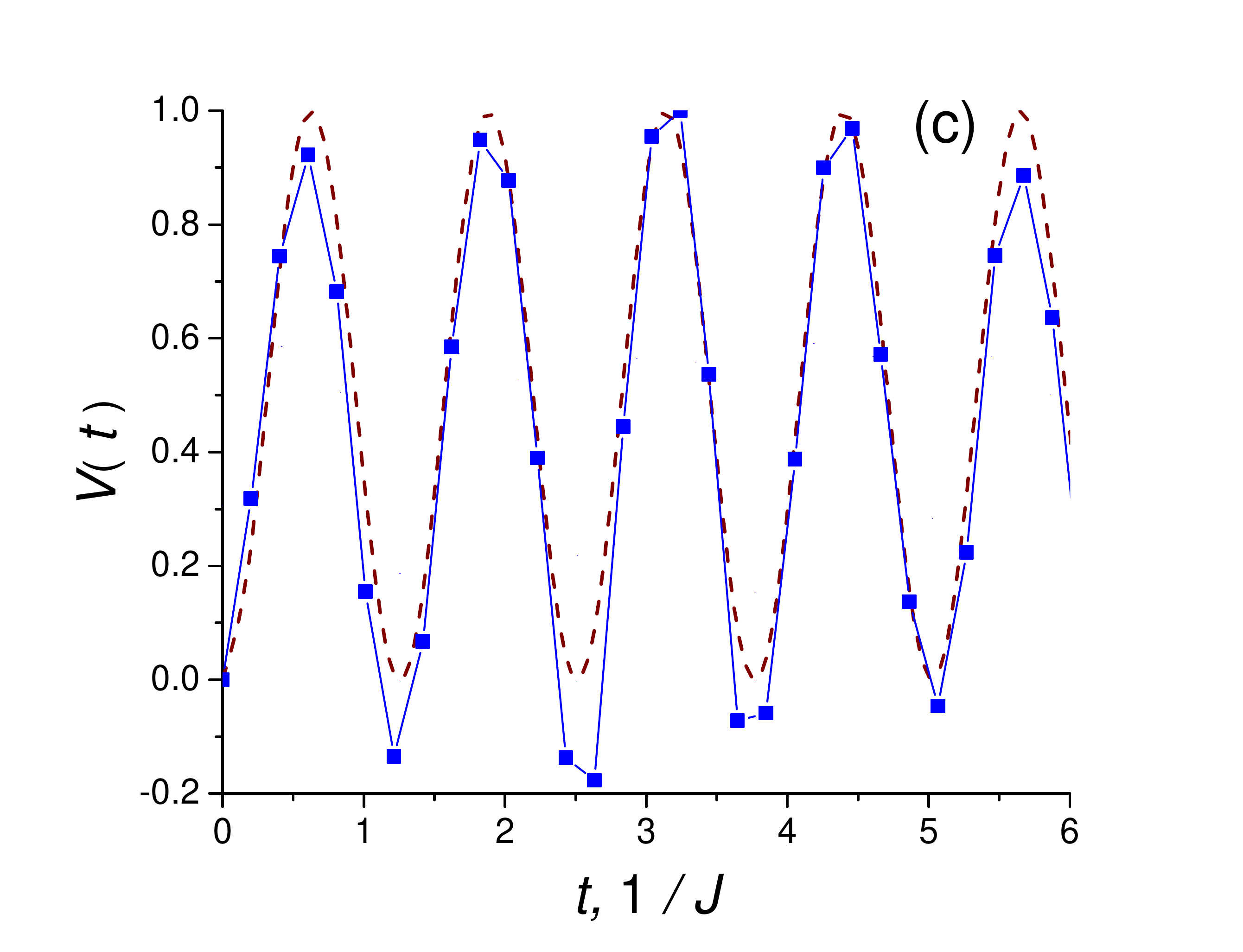}
    \caption{
        \label{Ising16N1}
        (Color online) The results of our experiment (solid blue lines) and theory (dashed brown lines) for $V$ defined in Eq. (\ref{Vtau}) in the case of the 16-spin transverse Ising ladder at $\alpha=J$ (a), $\alpha=2J$ (b), $\alpha=5J$ (c) as a function of the time $t$ for the Trotter number $N=1$. }
\end{figure}

\section{Summary and conclusions}

In this paper, we pointed out that superconducting quantum computers are perspective for the unitary simulation of far-from-equilibrium dynamics including quenches of various spin models in one and two dimensions. Spin models nowadays are considered as an important playground in studies of fundamental problems of statistical physics from the first principles, such as the phenomena of thermalization of closed quantum systems. The advantage of programmable quantum computers is that different spin models can be simulated algorithmically on the same chip. It is possible to tune the model from the integrable type to nonintegrable type and to study the impact of such a "crossover" on the evolution. Moreover, due to the individual addressability of physical qubits of the chip, one can create various initial conditions (different from Hamiltonian eigenstates) and to study the influence of these conditions on the free evolution of the system.

We here used 5-qubit and 16-qubit superconducting quantum computers of the IBM Quantum Experience to show concepts of such a digital modeling. Two different models have been addressed in our proof-of-principles experiments -- the integrable central spin model (5-qubit device) and the transverse-field Ising model in both the one-dimensional and ladder configurations (16-qubit device), which is integrable in the first case and nonintegrable in the second one. The choice of models is linked to the topologies of the quantum chips as well as to gate errors -- we have chosen spin models with the arrangement of spins and interactions between them, which are in one-to-one correspondence with the quantum chips and available two-qubit gates between physical qubits of the chips. We have shown that the quantum machines we used are able to reproduce some important aspects of system's dynamics originating from the character of the initial conditions. For instance, it is possible to simulate the effect of the excitation blockade which occurs due to the negative quantum interference of contributions from different spins under the antisymmetric choice of the initial entangled state of the system. However, the practical usability of the devices is limited by errors of two-qubit gates, so that no more than several Trotter steps in the decomposition of the evolution operator can be realized. Nevertheless, we applied some heuristic tricks in order to "reduce" the total error and to extract a valuable information on relative quantities from the experimental data.

Although the reported results can be relatively easily found explicitly or using classical computers, scaling towards chips with many physical qubits, improved coherence times and reduced CNOT errors might lead to the resolution of problems which can hardly be solved using more traditional approaches. Indeed, in order to study the dynamics from first principles, one needs to know all eigenstates of a given Hamiltonian. The number of eigenstates, in general, increases exponentially with the increase of the particle number. Therefore, even quantum computers of medium sizes, which can appear in the near future, might be of practical importance for the modeling of dynamics of quantum systems. Even if errors will be still too large to simulate the dynamics untractable on classical computers, an implementation of advanced error correction codes should ultimately allow for such a modeling.

\begin{acknowledgements}
    We acknowledge use of the IBM Quantum Experience for this
    work. The viewpoints expressed are those of the authors and
    do not reflect the official policy or position of IBM or the
    IBM Quantum Experience team.
    \end{acknowledgements}

\newpage
\appendix

\section{Preparation of three-particle entangled state}

\begin{figure}[htp]
\center
    \includegraphics[width=.75\linewidth]{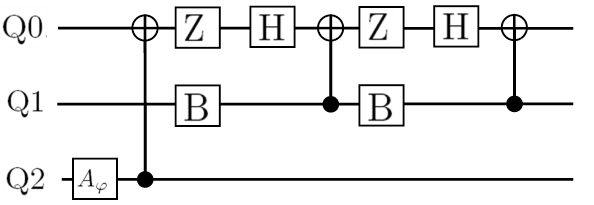}
    \caption{
        \label{3QES}
        Quantum circuit for the preparation of three-qubit excited state.}
\end{figure}

Quantum circuit used to prepare three-particle entangled state (\ref{init2}) is shown in Fig. \ref{3QES}. Single-qubit gates $A_{\varphi}$ and $B$ are constructed from the standard IBMqx4 gate $U_3$ as $A_{\varphi}=U_3(\theta = 2 \arccos \frac{1}{\sqrt{3}}, \varphi, \lambda = 0)$, $B=U_3(\theta = \frac{\pi}{4}, \varphi=0, \lambda = 0)$; Z is Pauli-Z gate.

\section{Full quantum circuits for the central spin model}

\begin{figure}[htp]
\center
    \includegraphics[width=0.75\linewidth]{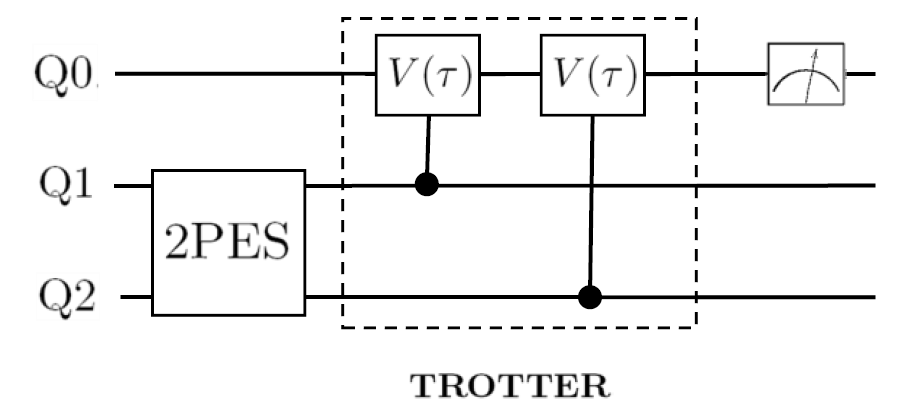}
    \caption{
        \label{fullcircuit1}
        Quantum circuit for the evolution of the system starting from the initial state of two-particle entangled state of the bath and unexcited central spin at the Trotter number $N=1$.}
\end{figure}

\begin{figure}[htp]
\center
    \includegraphics[width=0.75\linewidth]{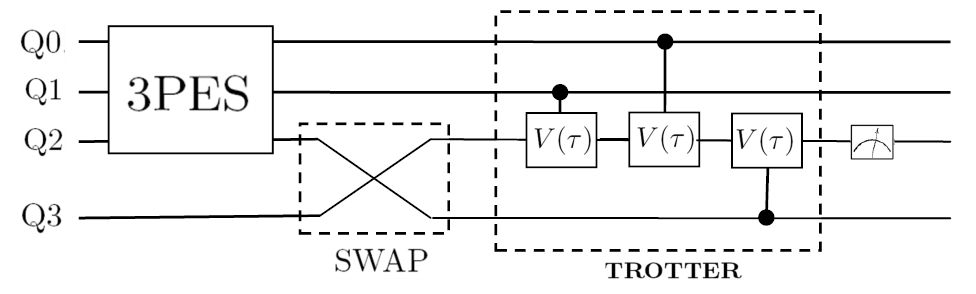}
    \caption{
        \label{fullcircuit2}
        Quantum circuit for the evolution of the system starting from the initial state of three-particle entangled state of the bath and unexcited central spin  at the Trotter number $N=1$.}
\end{figure}

\begin{figure}[htp]
\center
    \includegraphics[width=0.75\linewidth]{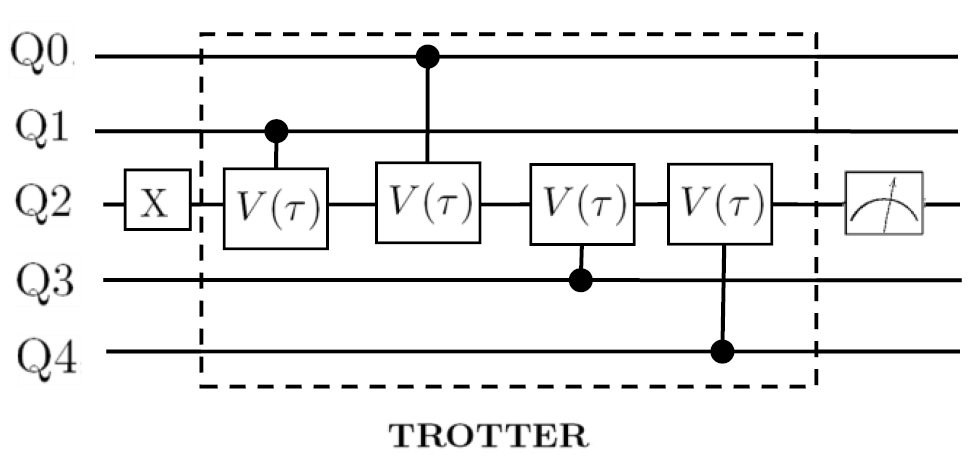}
    \caption{
        \label{fullcircuit3}
        Quantum circuit for the evolution of the system starting from the initial state of excited central spin and four unexcited spins of the bath  at the Trotter number $N=1$.}
\end{figure}

Figures \ref{fullcircuit1}, \ref{fullcircuit2}, and \ref{fullcircuit3} show full quantum circuits for three different situation within the central spin Hamiltonian. For the sake of simplicity, we restrict ourselves to the single Trotter number, $N=1$. The generalization to the circuits with $N > 1$ is straightforward.

\FloatBarrier

\section{Raw results for the 8-spin transverse-field Ising chain}

\begin{figure}[h]
\center
    \includegraphics[width=.48\linewidth]{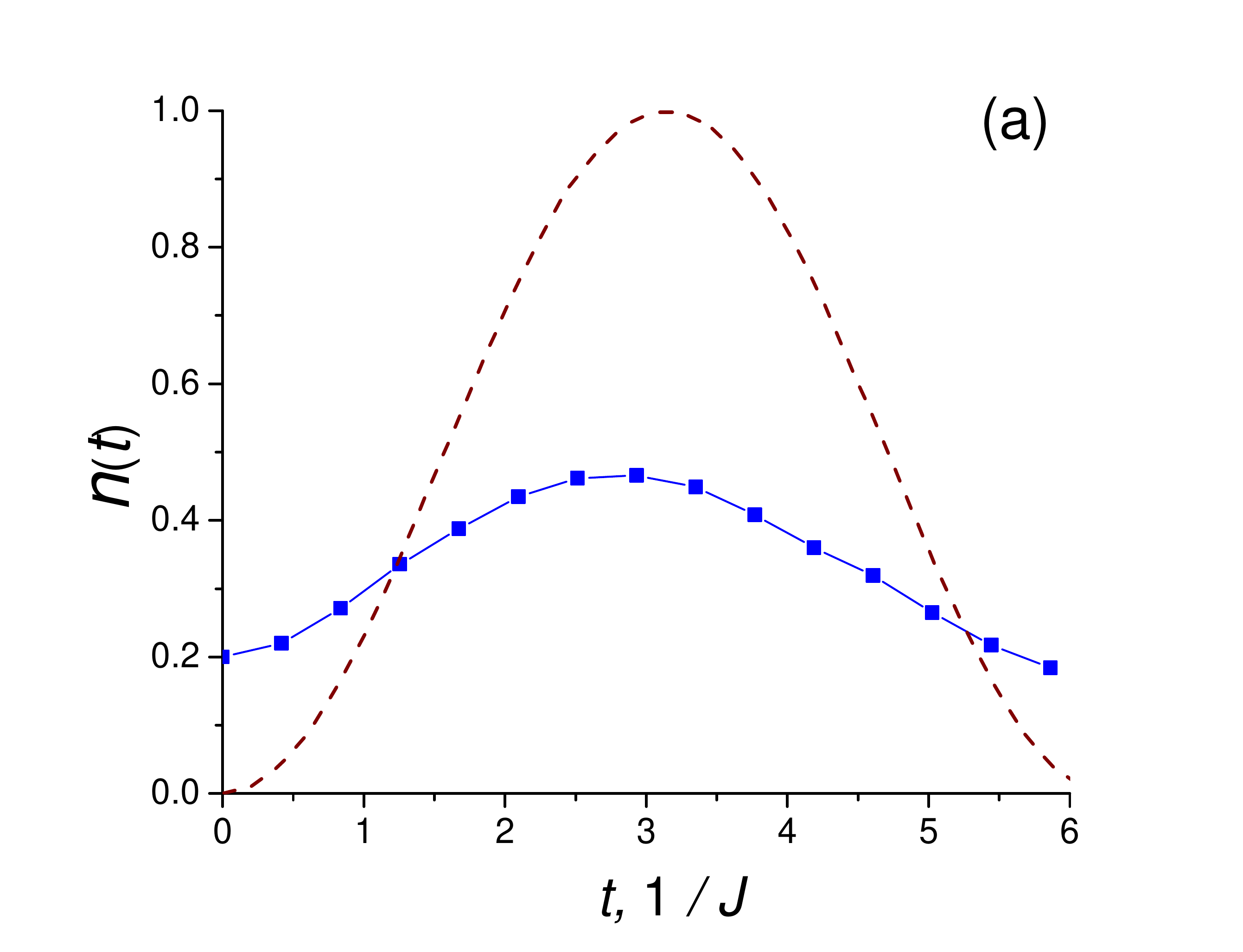}
    \includegraphics[width=.48\linewidth]{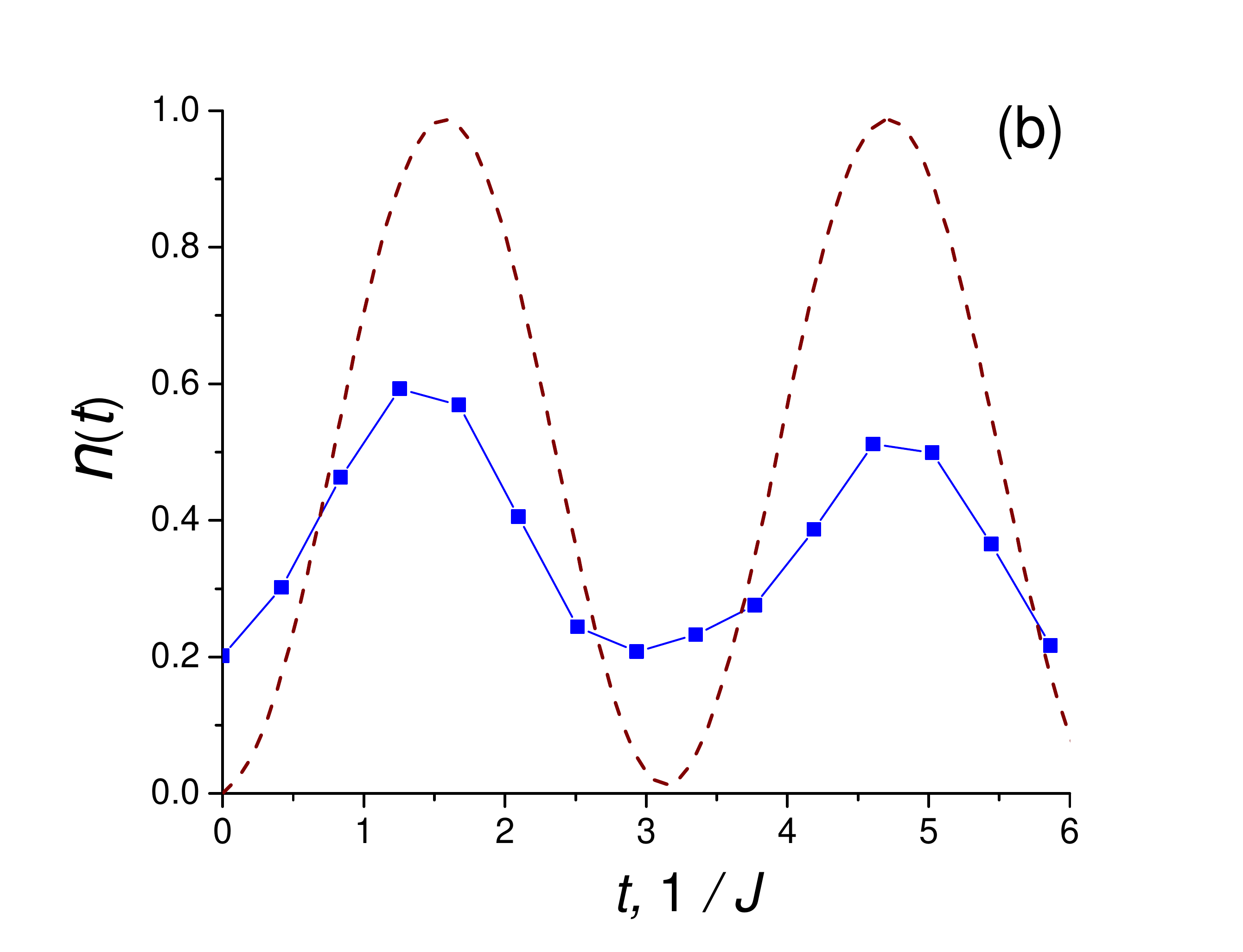}
    \includegraphics[width=.48\linewidth]{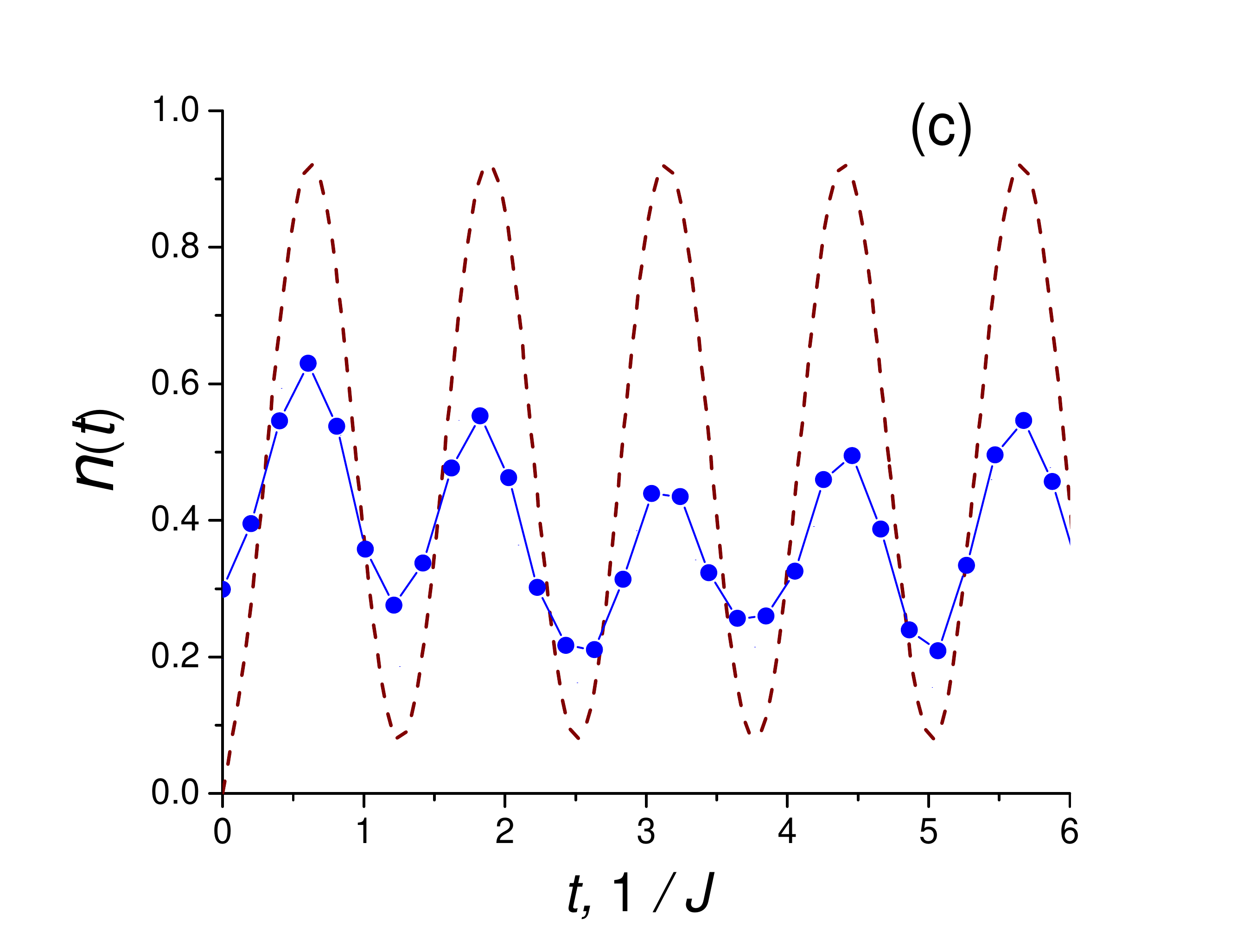}
    \caption{
        \label{appIsing8N2}
        (Color online) The results of our experiment (solid blue lines) and theory (dashed brown lines) for the occupations of the upper levels of 8-spin transverse Ising chain at $\alpha=J$ (a), $\alpha=2J$ (b), $\alpha=5J$ (c) as a function of the time $t$ for the Trotter number $N=2$. }
\end{figure}

Figure \ref{appIsing8N2} provides raw data for the occupations of the upper levels of the 8-spin transverse-field Ising chain simulated in the real device in comparison with the theoretical results. In both cases, Trotter number is $N=2$.

\FloatBarrier
\section{Raw results for the 16-spin transverse-field Ising ladder}

\begin{figure}[htp]
\center
    \includegraphics[width=.48\linewidth]{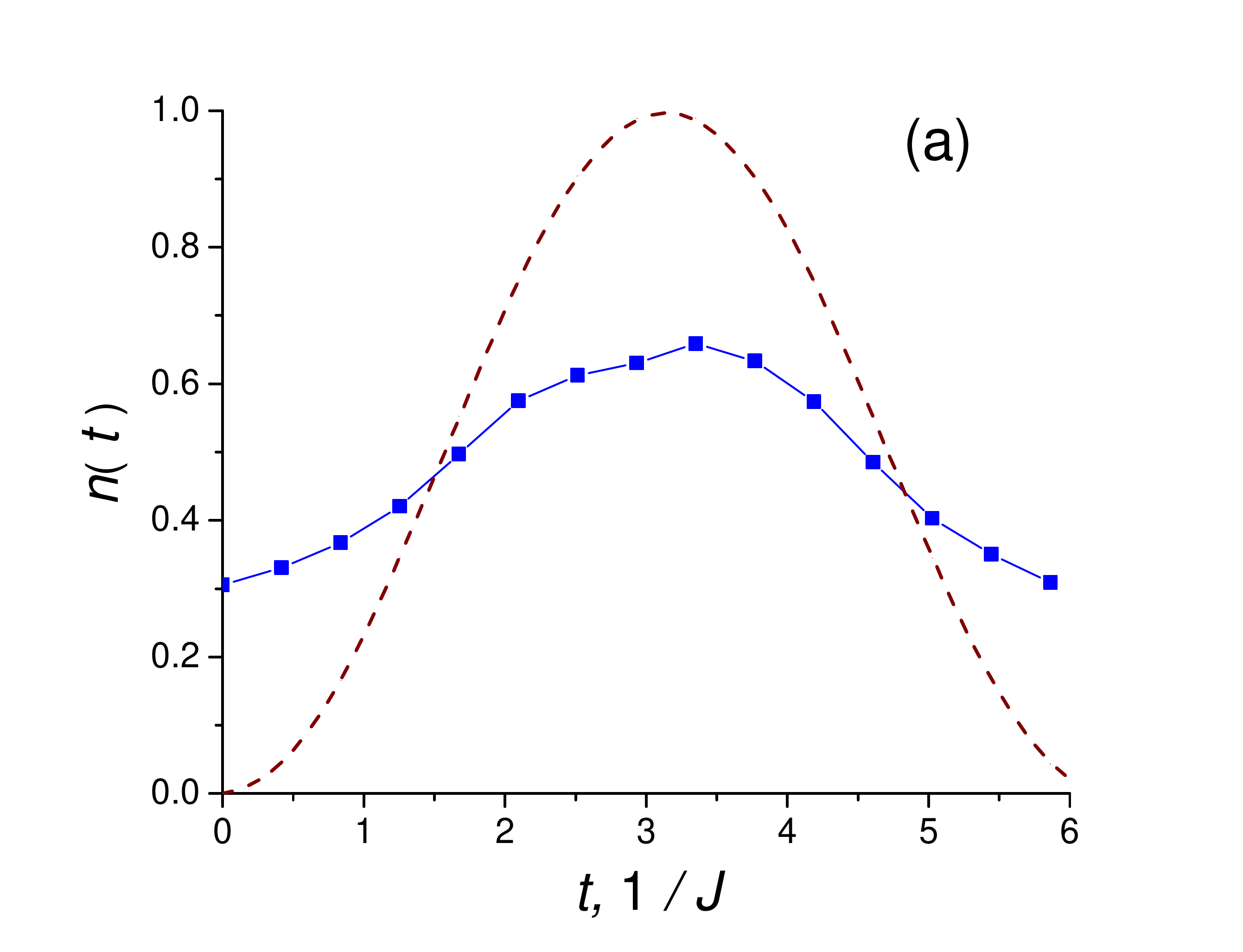}
    \includegraphics[width=.48\linewidth]{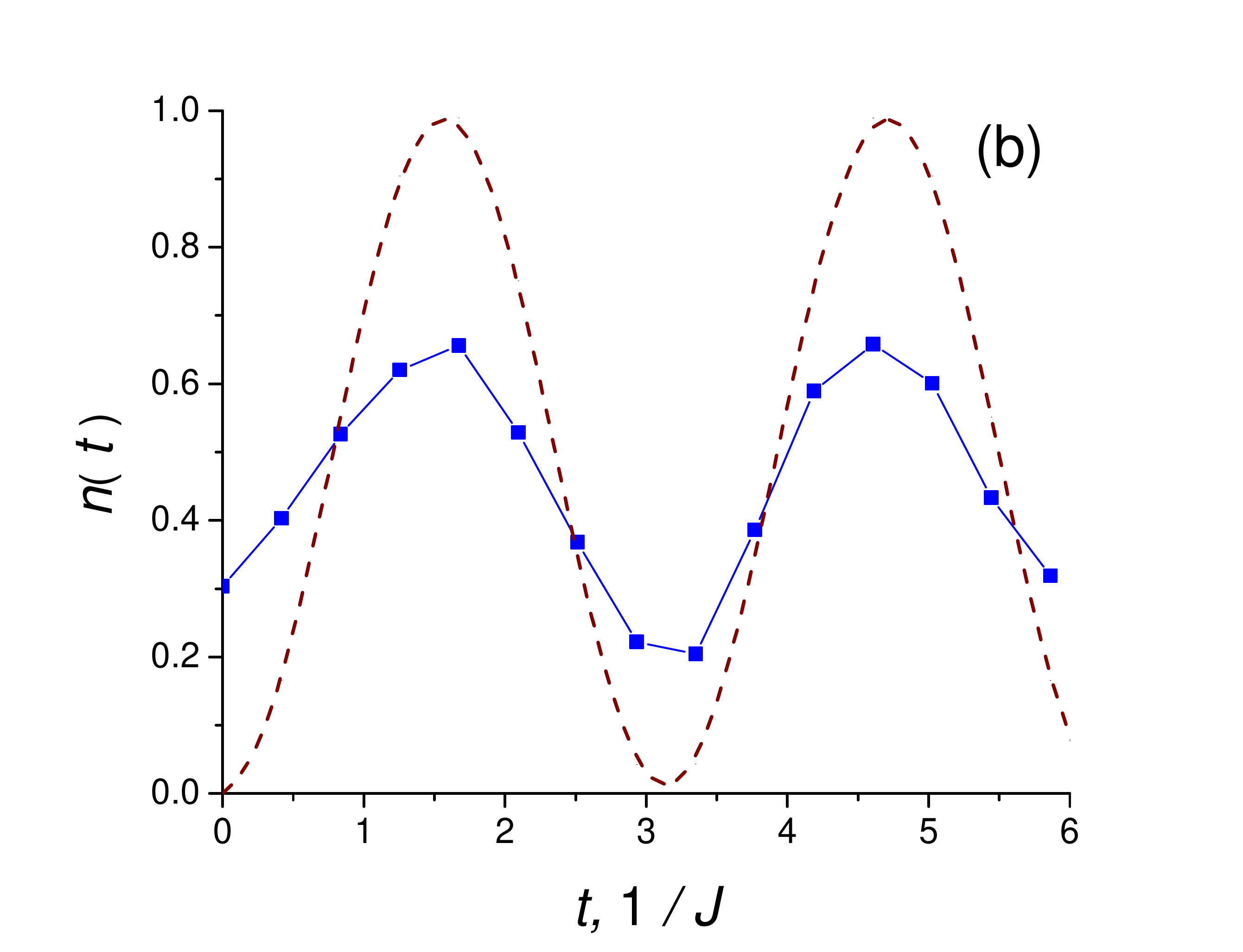}
    \includegraphics[width=.48\linewidth]{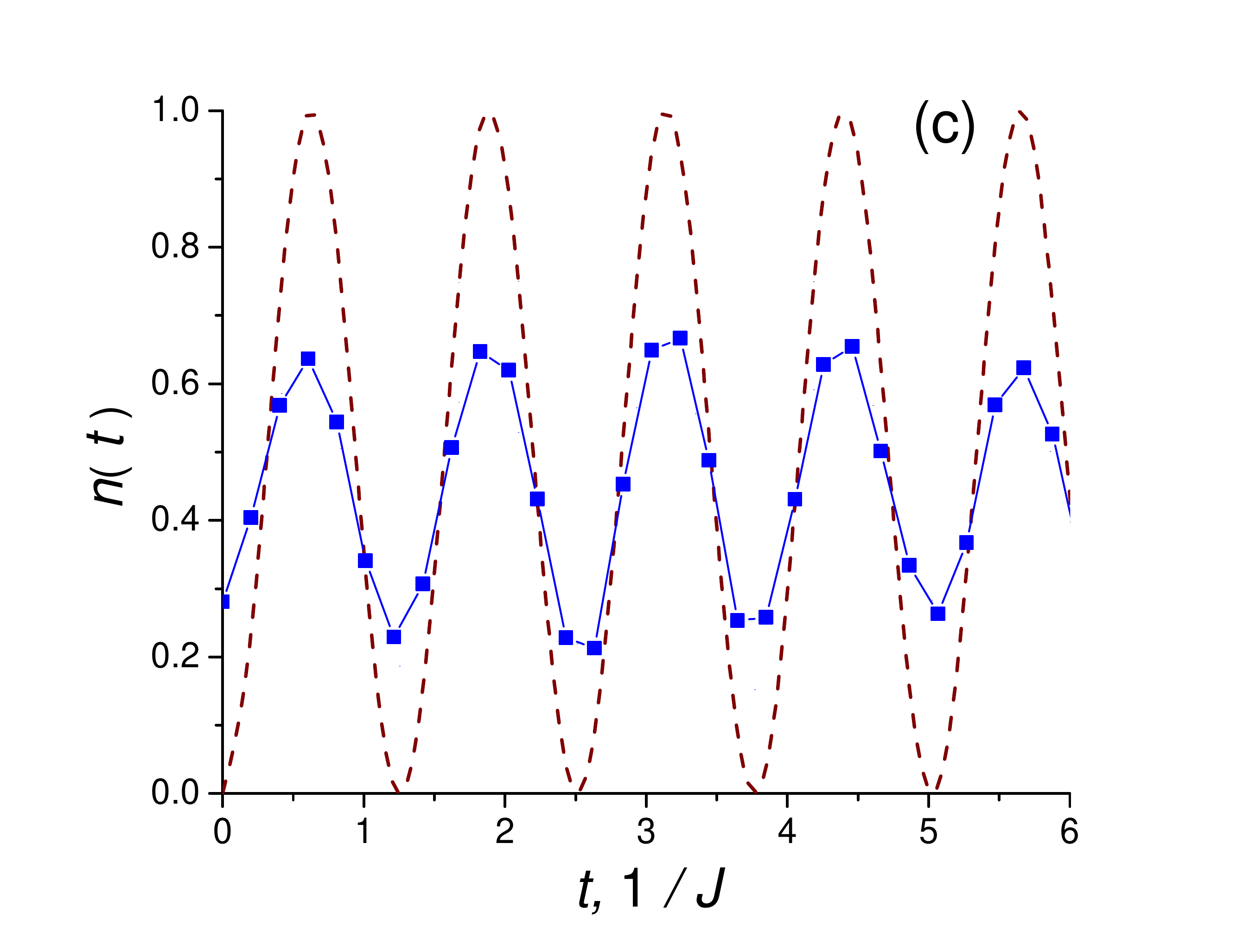}
    \caption{
        \label{appIsing16N1}
        (Color online) The results of our experiment (solid blue lines) and theory (dashed brown lines) for the occupations of the upper levels of 16-spin transverse Ising ladder at $\alpha=J$ (a), $\alpha=2J$ (b), $\alpha=5J$ (c) as a function of the time $t$ for the Trotter number $N=1$. }
\end{figure}

Figure \ref{appIsing16N1} provides raw data for the occupations of the upper levels of 16-spin transverse-field Ising ladder simulated in the real device in comparison with the theoretical results. In both cases, Trotter number is $N=1$.

\FloatBarrier

\end{document}